\documentclass[useAMS,usenatbib,usegraphicx]{mn2e}  
\usepackage[usenames]{color}

\newcommand{\be}{\begin{equation}}
\newcommand{\ee}{\end{equation}}
\newcommand{\bea}{\begin{eqnarray}}
\newcommand{\eea}{\end{eqnarray}}

 \title[DM in AS1063] 
{A Free-Form mass model of the Hubble Frontier Fields Cluster AS1063 (RXC J2248.7-4431) with over one hundred constraints.} 
\author[J.M. Diego]  
  {Jose M. Diego\thanks{jdiego@ifca.unican.es}$^1$, Tom Broadhurst$^{2,3}$, Jess Wong$^4$,  Joseph Silk$^{6,7,8,9}$, Jeremy Lim$^{4,5}$,
  \newauthor
  Wei Zheng$^8$, Daniel Lam$^4$, Holland Ford$^8$\\
$^{1}$IFCA, Instituto de F\'isica de Cantabria (UC-CSIC), Av. de Los Castros s/n, 39005 Santander, Spain\\
$^{2}$  Fisika Teorikoa, Zientzia eta Teknologia Fakultatea, Euskal Herriko Unibertsitatea UPV/EHU, E-48080 Bilbao, Spain\\   
$^3$IKERBASQUE, Basque Foundation for Science, Alameda Urquijo, 36-5 48008 Bilbao, Spain\\
$^4$ Department of Physics, The University of Hong Kong, 0000-0002-6536-5575, Pokfulam Road, Hong Kong\\
$^5$ Laboratory for Space Research, Faculty of Science, The University of Hong Kong, 0000-0002-6536-5575, Pokfulam Road, Hong Kong\\
$^6$ Institut d\'Astrophysique de Paris (UMR 7095: CNRS \& UPMC - Sorbonne Universités), 98 bis bd Arago, F-75014 Paris, France\\
$^7$ Laboratoire AIM-Paris-Saclay, CEA/DSM/IRFU, CNRS, Univ. Paris VII, F-91191 Gif-sur-Yvette, France\\
$^8$ Department of Physics and Astronomy, The Johns Hopkins University Homewood Campus, Baltimore, MD 21218, USA\\ 
$^9$ BIPAC, Department of Physics, University of Oxford, Keble Road, Oxford OX1 3RH, UK
}

\date{Draft version \today}  
  
\pagerange{\pageref{firstpage}--\pageref{lastpage}}  
  
\begin{document}  
\maketitle  
 
\label{firstpage}  
\begin{abstract} 

We derive a free-form mass distribution for the massive cluster AS1063 (z=0.348) using the completed optical imaging from the Hubble Frontier Fields programme. Based on a subset of 11 multiply lensed systems with spectroscopic redshift we produce a lens model that is accurate enough to unveil new multiply lensed systems, totalling over a 100 arclets,  and to estimate their redshifts geometrically. Consistency is found between this precise model and that obtained using only the subset of lensed sources with spectroscopically measured redshifts.Although a relatively large elongation of the mass distribution is apparent relative to the X-ray map, no significant offset is found between the centroid of our mass distribution and that of the X-ray emission map, suggesting a relatively relaxed state for this cluster. For the well resolved lensed images we provide detailed model comparisons to illustrate the precision of our model and hence the reliability of our de-lensed sources. A clear linear structure is associated with one such source extending ~23 kpc in length, that could be an example of jet-induced star formation, at redshift $z \approx 3.1$.

\end{abstract}  
  
\section{Introduction}\label{sect_intro}  
The Hubble Frontier Fields (HFF) program\footnote{http://www.stsci.edu/hst/campaigns/frontier-fields/} \citep{Lotz2014} provides deep observations of the distant Universe through 
some of the most spectacular natural cosmic lenses. The HFF programme makes it possible to study the mass distribution in the central region of merging clusters in detail through  
the lensing distortion induced in background galaxies that results in typically over a hundred multiply lensed images. Most of these images are faint and small making it hard to identify sets of counter images without the guidance of a reliable model. The HFF clusters are chosen on the basis of having the largest known Einstein radii, comprising particularly massive clusters with obvious ongoing interaction. These clusters have complex critical curves that are far from symmetric, compounding the problem of identifying multiple images. During a major merger the critical curves can be {\it stretched} between the mass components enhancing the critical area with elongated critical curves \citep{Redlich2012}. An extreme example is the cluster MACS0717 \citep{Diego2015b} where as many as 4 massive clusters are merging, producing a critical area stretching  over 2 arc minutes in length, subjected to very large magnification. 

In this paper, we explore the cluster AS1063 \citep[z=0.348][also known as RXC J2248.7-4431]{Guzzo2009}. This cluster is currently being observed within the HFF programme. The collection of optical data in the central part of this cluster ha been completed permitting here a significant improvement in resolution of the mass map and associated magnification field, needed for understanding this complex cluster and the nature of the magnified background galaxies. 
AS1063  is one of the hottest X-ray clusters possibly undergoing a major merger \citep{Gomez2012}. 
This cluster has been studied previously in the context of gravitational lensing because of its status as one of the largest lenses in the Southern sky. 
It was chosen as a target for the CLASH program, which confirmed it as a valuable lens by virtue of its large magnification\citep{Balestra2013,Boone2013,Gruen2013,Bradley2014,Johnson2014,Monna2014,Richard2014,Umetsu2015,Zitrin2015}. Recently, VLT spectroscopy has provided accurate redshifts for many of the  previously known arclets \cite{Caminha2015a}.

To date, $\sim 20$ multiply lensed galaxies have been reliably identified  \citep{Johnson2014,Richard2014,Caminha2015a} (see also the recent results from GLASS), 
in the redshift range $1<z<6$ \citep{Boone2013,Bradley2014,Monna2014} in the cluster field. There is general agreement that the mass distribution has a symmetric relaxed form with an obvious large-scale elongation \citep{Richard2014}. This large-scale elongation is aligned with the major axis of a prominent central BCG \citep{Johnson2014,Monna2014,Richard2014,Zitrin2015}, having a similar elongation. This elongation extends well beyond the virial radius of the cluster \citep{Gruen2013}, but with an azimuthally averaged large scale massive profile that fits well the NFW form out to a radius of 2 Mpc/h \citep{Umetsu2015}. In this paper, we make a strong lensing analysis of the deep new optical HFF data now completed for this cluster, using our general  free-form lensing technique \citep{Diego2005,Diego2007,Diego2016,Sendra2014}.  Our aim is to objectively identify new multiply lensed systems for understanding the properties of low luminosity galaxies in the high redshift Universe.  

To date, the HFF program has revealed unprecedented numbers of multiply lensed galaxies reaching a limit of z$\simeq$ 9.6 \citep{Zitrin2014}. Interestingly, despite 2 magnitudes of magnification and the great depth of this imaging, no galaxy has yet been discovered beyond the  most distant galaxies already known at $z \sim 10$ \citep{Coe2013,Zitrin2014,Zheng2014,Oesch2014,Coe2015,Ishigaki2015}. This difficulty is not 
due to the filter choice, which in principle can access Lyman-break galaxies out to z$\simeq$ 12.0.  As the HFF program progresses to completion we may anticipate a clear, field averaged constraint on the number density of galaxies lying above $z>9.0$. A significant absence of such galaxies is not predicted for the LCDM model, where several galaxies are expected per HFF cluster in the range $9<z<10$ on the basis of the standard LCDM model, by extrapolating the luminosity function \citep[e.g][]{Coe2015,Schive2015}. This may have  profound implications for the nature of dark matter as this absence at $z>9.0$ is a distinct prediction of the wave-DM scenario where light bosons such as axions lie in a ground state, and for which the inherent Jeans scale in this context suppresses the formation of low mass galaxies thereby delaying galaxy formation relative to LCDM \citep{Schive2014,Schive2015,Bozek2015}. This scenario has only one free parameter, the boson mass, which is constrained to be $\simeq 10^{-22}$eV by the local dwarf cores and which translates into a sharp onset of galaxy formation at $z \sim 9-10$  \citep{Bozek2015,Schive2014,Schive2015}. This precise prediction means this model is readily falsifiable if significant numbers of galaxies were to be found at $z>10$, and hence the new constraints provided by the HFF in this redshift regime provides a very crucial, timely means of discrimination between the wave-DM model and standard heavy particle interpretation of CDM.

\begin{figure}    
 {\includegraphics[width=8.5cm]{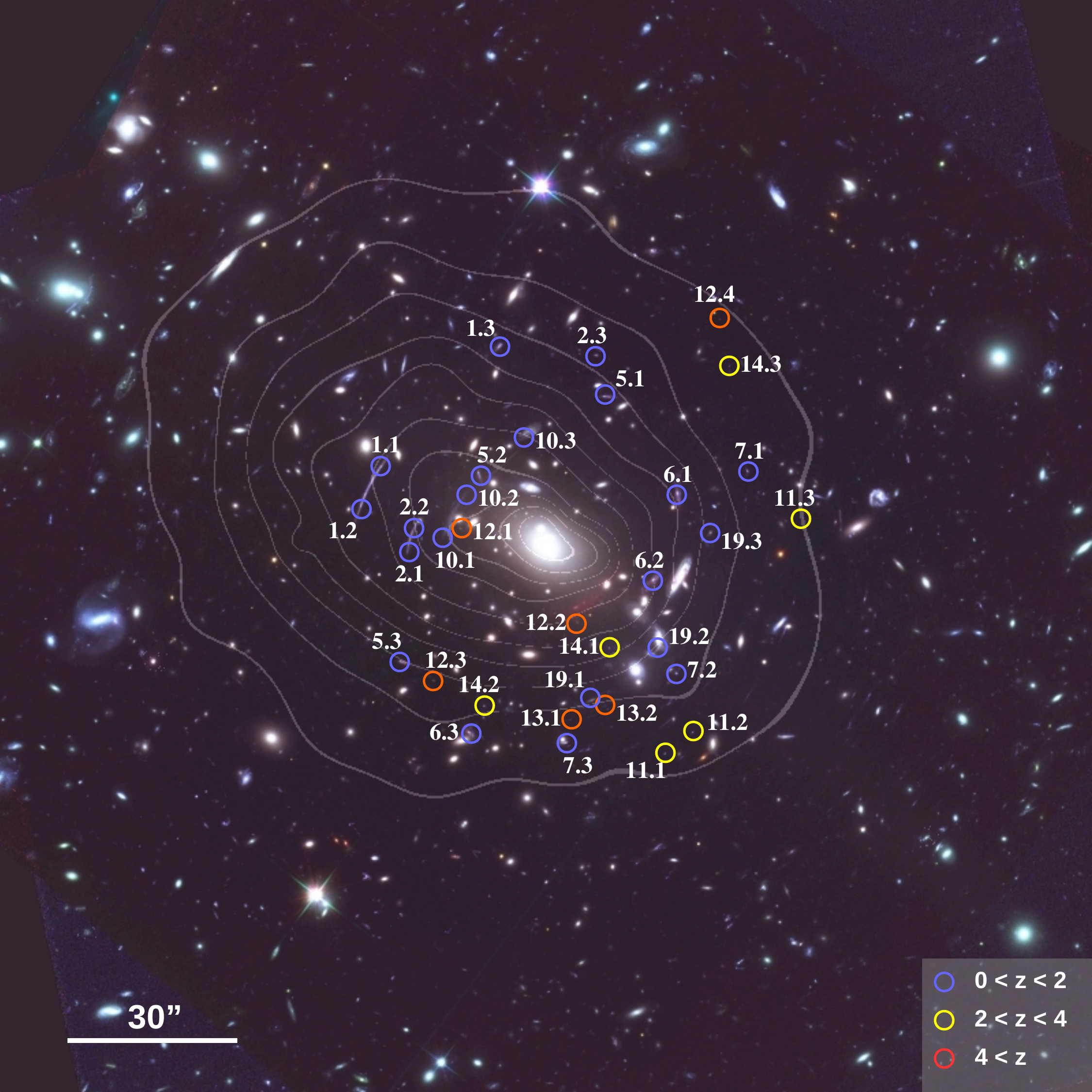}}
   \caption{AS1063 as seen by HST with {\it Chandra} contours overlaid on top. The field of view is 3.2\arcmin. The circles mark the positions of the multiply 
            lensed systems with spectroscopic redshift that are used to build the preliminary (or driver) lens model.}
   \label{fig_AS1063_Chandra}  
\end{figure}  

This paper is organised as follows. In section~\ref{sect_S2} we introduce the HFF data used in this study and briefly describe the X-ray data. 
In section~\ref{sect_S4} we present the initial lensing data used to constrain our preliminary model. Section~\ref{sect_S5} describes the algorithm used to derive the lens models. The results are presented in section~\ref{sect_S7} and they are discussed in section~\ref{Sect_Discus}. 

Throughout the paper we assume a cosmological model with $\Omega_M=0.3$,
$\Lambda=0.7$, $h=70$ km/s/Mpc. For this model, 1\arcsec equals $4.92$ kpc at the distance of the cluster.

\section{HFF data and X-ray data}\label{sect_S2}

We used public imaging data obtained from the ACS and WFC3 Hubble instruments, retrieved from the Mikulski Archive for Space 
Telescope (MAST). For the optical data (filters: F435W, F606W and F814W), we used the recently released data that includes the first 
50 orbits of HFF data on this cluster  (ID 14037, PI. J. Lotz) plus 4 orbits from the previous CLASH programme \citep{Postman2012} 
(ID 12458, PI. M. Postman). For the IR data, we used data collected in previous programmes in the filters F105W (2 orbits),F125W (2 orbits), 
F140W (2 orbits), and F160W (2 orbits), (IDs 12458, PI M. Postman and 13459 PI T. Treu). From the original files, we produce two sets of color images by combining the optical and IR bands. The first set is based on the raw data while in the second set we apply a 
high-pass filter to reduce the diffuse emission from member galaxies and a high-pass filter to increase the signal-to-noise ratio of small compact faint objects. The second set is particularly useful to match colors in objects that lie behind a luminous member galaxy where the light from the foreground galaxy affects the colors  of the background galaxy. 

To explore the dynamical state of this cluster, we produced an X-ray image using public Chandra data on this cluster. 
In particular, we used data with the Obs ID 4966 (PI. Romer) totaling 26.7 ks. 
The X-ray data is smoothed using the code {\small ASMOOTH} \citep{Ebeling2006}. 
Both the HFF and the smoothed X-ray map are shown in figure \ref{fig_AS1063_Chandra}. 
No  offset is observed between the peak of the X-ray emission and the BCG. The BCG itself shows no excess X-rays with respect 
to the surrounding emission. The X-ray emission shows a clear elongation in the diagonal direction.

\section{Lensing data}\label{sect_S4}

For the lensing data we follow the recent multiple-image system identifications from \cite{Johnson2014} and \cite{Richard2014} that 
include 19 multiply lensed systems (see compilation in Table~\ref{tab1} below). 
From these papers we also adopt their numbering system as well as their spectroscopic redshifts. 
We also use the new spectroscopic redshifts from \cite{Caminha2015a}, providing new independent spectrospcopic redshifts of previously established multiply lensed systems and new spectroscopic redshifts of the lensed systems 7 and 14 (bringing the total number of systems with spectroscopic redshift to 11) that were also previously known but had no spectroscopic redshifts. 
In \cite{Karman2015}, new redshifts for three of the multiple image families (13,19,52) are given and confirms the multiple images of several others.
Although not used in this work, at the time of writing this manuscript, additional data from GLASS\footnote{https://archive.stsci.edu/prepds/glass/} \citep{Treu2015} was released providing useful redshift information of cluster members and background sources. Amog them, they confirmed the redshift of at least one of the multiple lensed images (system 10).  
Among the new redshifts, the spectroscopic redshift of our new system 10 is of particular interest as this system has not hitherto been recognized as a multiply-lensed system (we identify the counterimage in this work). The spectroscopic redshift agrees remarkably well with the geometric redshift inferred (before we knew about the spectroscopic redshift) from our initial lens model based on all previously known systems, listed in Table~\ref{tab1}\footnote{Support material including footstamps of the entries in Table~\ref{tab1} can be found in http://www.ifca.unican.es/users/jdiego/AS1063}. This system, together with the agreement between the new spectroscopic redshifts and the those we derive in a blind way from our lens model are discussed in more detail in section~\ref{Sect_Syst10}. The resulting subset of 11 systems with spectroscopic redshifts is shown in Fig.~\ref{fig_AS1063_Chandra}. The positions and redshifts of all systems used in this paper are listed in Table~\ref{tab1} in the appendix. 

In addition to the centroid position of the multiply lensed systems, we can also use the position of individual knots that are readily identified in the different  counterimages thanks to the depth of the HFF data. In particular, systems 1, 2, 5, 6, 10 and 19 contain distinguishing features that can be easily identified in the multiple images.  In the context of our free-form model method, the addition of extra knots in well resolved systems greatly improves the accuracy and stability of the derived lensing solutions \citep{Diego2016}. In a later iteration of the reconstruction (see section \ref{sect_S7} for details on the iterative process), we include also pixels tracing the length of conspicuous elongated arcs (that are not necessarily multiply lensed although some of them may be) as additional constraints by requiring that these arcs have to focus to a small region in the source plane. This additional information is especially useful in the regions beyond the critical curves where the density of constraints drops.

Among the systems identified in the literature, system 12 in Table~\ref{tab1} contains 5 counterimages \citep{Balestra2013,Boone2013,Bradley2014,Monna2014,Caminha2015a}.  
One of the images is close to the BCG and although possibly a real counterimage, we do not use it as a constraint bacause   
other possible images exist in the vicinity of the central image 12.5 predicted in this vicinity. As discussed later, however, an a posteriori comparison of the predicted position of 
12.5 and the observed candidate identified in previous work does show a very good match with an estimated error of $\approx 1''$, see Fig.~\ref{fig_Centre}). 

\begin{figure}    
 {\includegraphics[width=8.5cm]{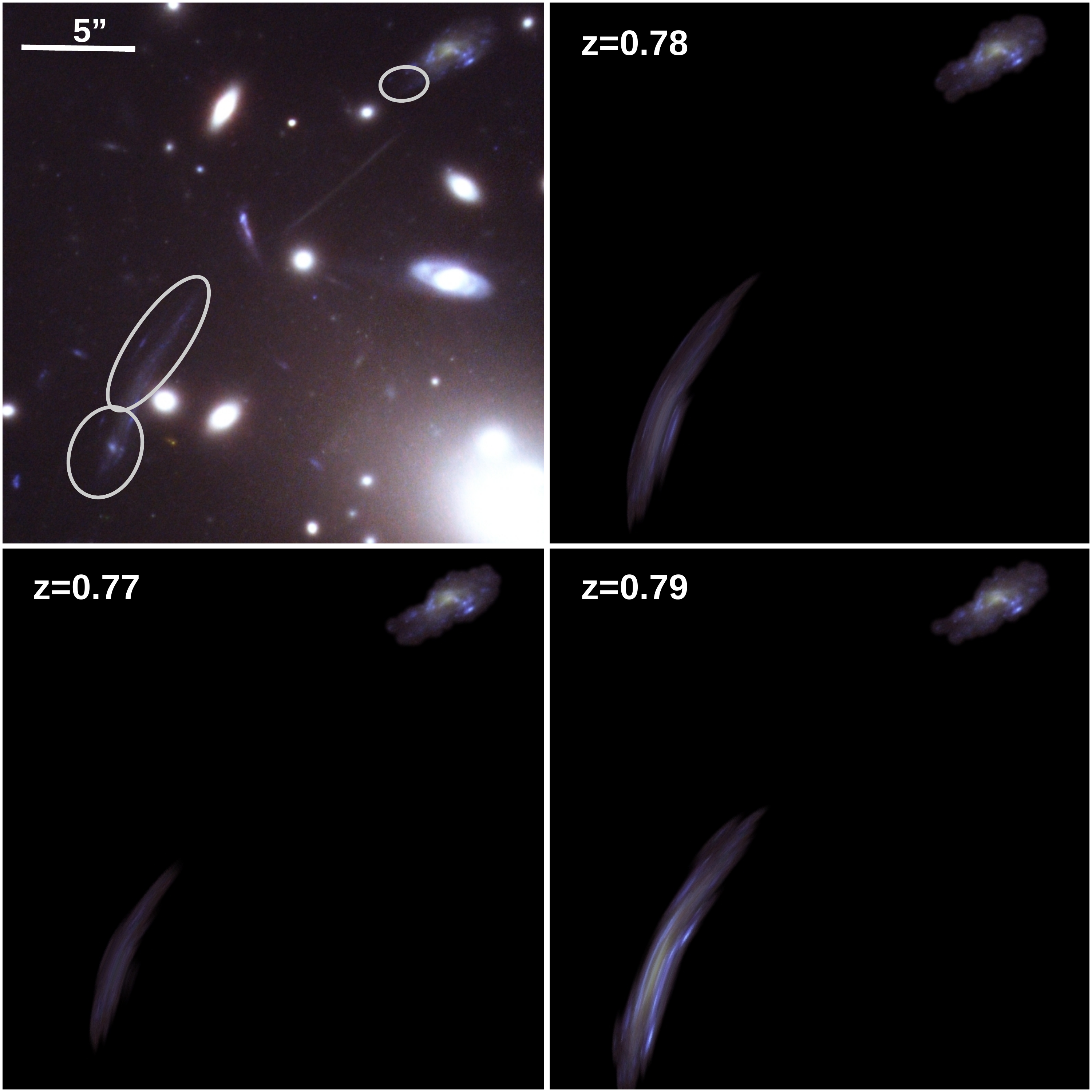}}
   \caption{Redshift prediction for our new system 10 when system 10 is not used in the derivation of the lens model. The ellipses mark the region that is multiply lensed. 
            The observed arc is better reproduced when z=0.78.  
            Changing this redshift by $\sim 0.01$ results in significant differences like the presence of an unobserved nucleus or the complete disappearance of the arc.}
   \label{fig_Syst10}  
\end{figure}  

\begin{figure}  
 \centerline{ \includegraphics[width=8.5cm]{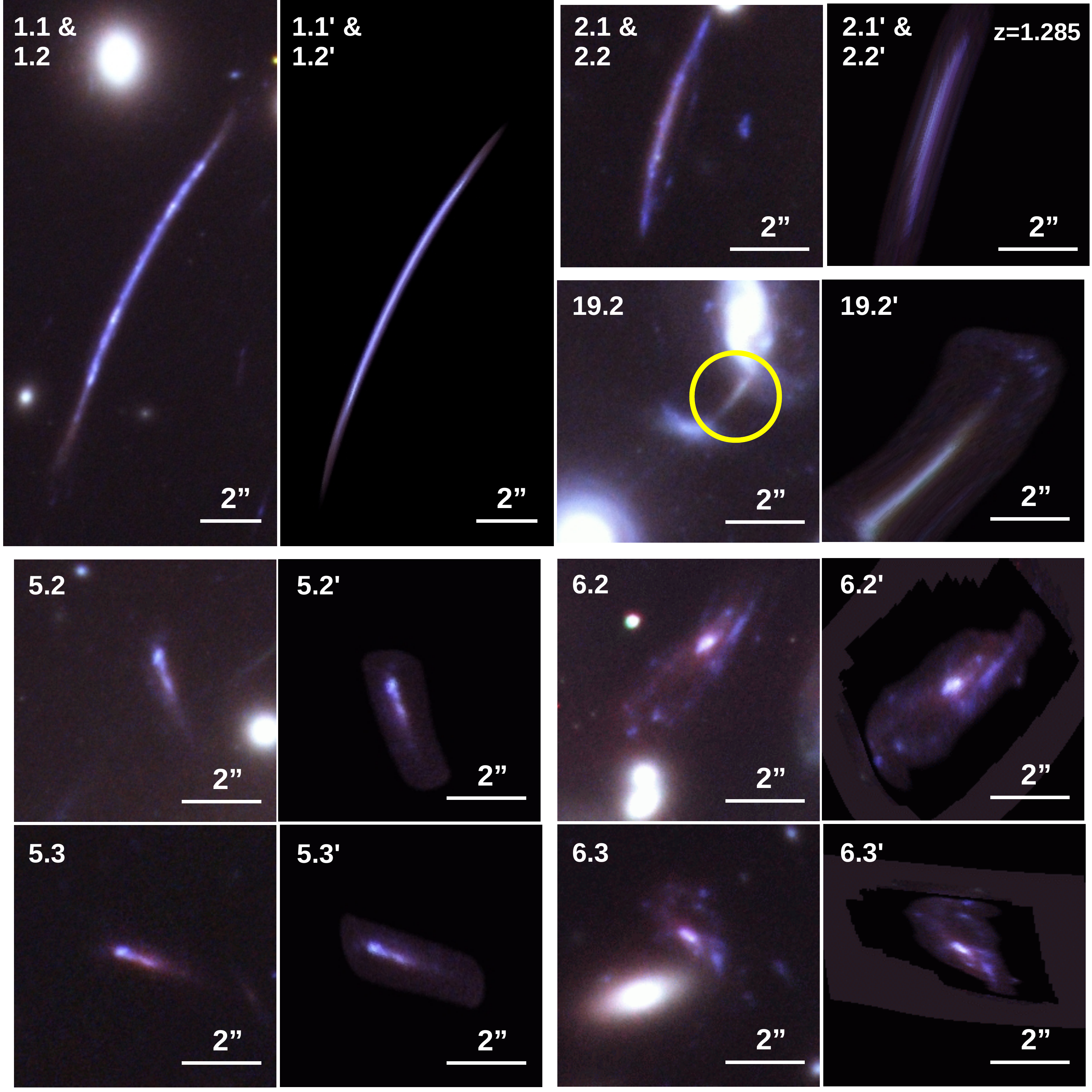}}  
   \caption{Relensed images for some resolved systems. For each image we show the original data and the model marked with a prime symbol ('). 
            We use the counterimage not shown in this figure to predict the other counterimage(s). For instance, 5.2' is predicted from 5.1 etc. 
            In all cases, both data and model are centered in the exact same coordinates. All models agree well with the data except for 
            19.2' (marked with a yellow circle) which is is affected by the presence of a nearby spiral galaxy (see text). 
            For system 2, the redshift of the system has been increased by $\approx 2\%$ in order to reproduce the observed arc (see text for a discussion).} 
   \label{fig_Relensed}  
\end{figure}  
 
\section{Lensing reconstruction algorithm; WSLAP+}\label{sect_S5}
We use our method WSLAP+ to perform the lensing mass reconstruction
with the lensed systems and internal features described above.
The reader can find the details of the method in our previous papers
\citep{Diego2005,Diego2007,Diego2016,Sendra2014}. 
Here we give a brief summary of the most essential elements. \\
Given the standard lens equation, 
\begin{equation} \beta = \theta -
\alpha(\theta,\Sigma), 
\label{eq_lens} 
\end{equation} 
where $\theta$ is the observed position of the source, $\alpha$ is the
deflection angle, $\Sigma(\theta)$ is the surface mass density of the
cluster at the position $\theta$, and $\beta$ is the position of
the background source. Both the strong lensing and weak lensing
observables can be expressed in terms of derivatives of the lensing
potential. 
\begin{equation}
\label{2-dim_potential} 
\psi(\theta) = \frac{4 G D_{l}D_{ls}}{c^2 D_{s}} \int d^2\theta'
\Sigma(\theta')ln(|\theta - \theta'|), \label{eq_psi} 
\end{equation}
where $D_l$, $D_s$, and $D_{ls}$ are the
angular diameter distances to the lens, to the source and from the lens to 
the source, respectively. The unknowns of the lensing
problem are in general the surface mass density and the positions of
the background sources in the source plane. 
The surface mass density is described by the combination of two components; 
i) a soft (or diffuse) component (parametrized as superposition of Gaussians) and 
ii) a compact component that accounts for the mass associated with the individual halos (galaxies) in the cluster. \\
For the diffuse component other functions could be used instead of Gaussians but the 
Gaussian functions provide a good compromise between the desired compactness and smoothness  of the basis function.
For the compact component we adopt directly the light distribution in one of the bands (F814W). To each galaxy, we assign an arbitrary mass 
proportional to its surface brightness. This mass is later re-adjusted as part of the optimization process. 
Alternatively, in previous works we have also considered NFW profiles associated to each member galaxy. 
The choice of either NFW or observed surface brightness plays a secondary role as shown in our earlier work 
\citep[see for instance][]{Diego2015a}. The compact component is usually divided 
in independent layers, each one containing one or several cluster members. The separation into different layers allows us to 
constrain the mass associated to special halos (such as the giant elliptical galaxies) independently from more 
ordinary galaxies. This is useful in the case where the light-to-mass ratio may be different, like for instance in the BCG. \\

\begin{figure*}  
 \centerline{ \includegraphics[width=16cm]{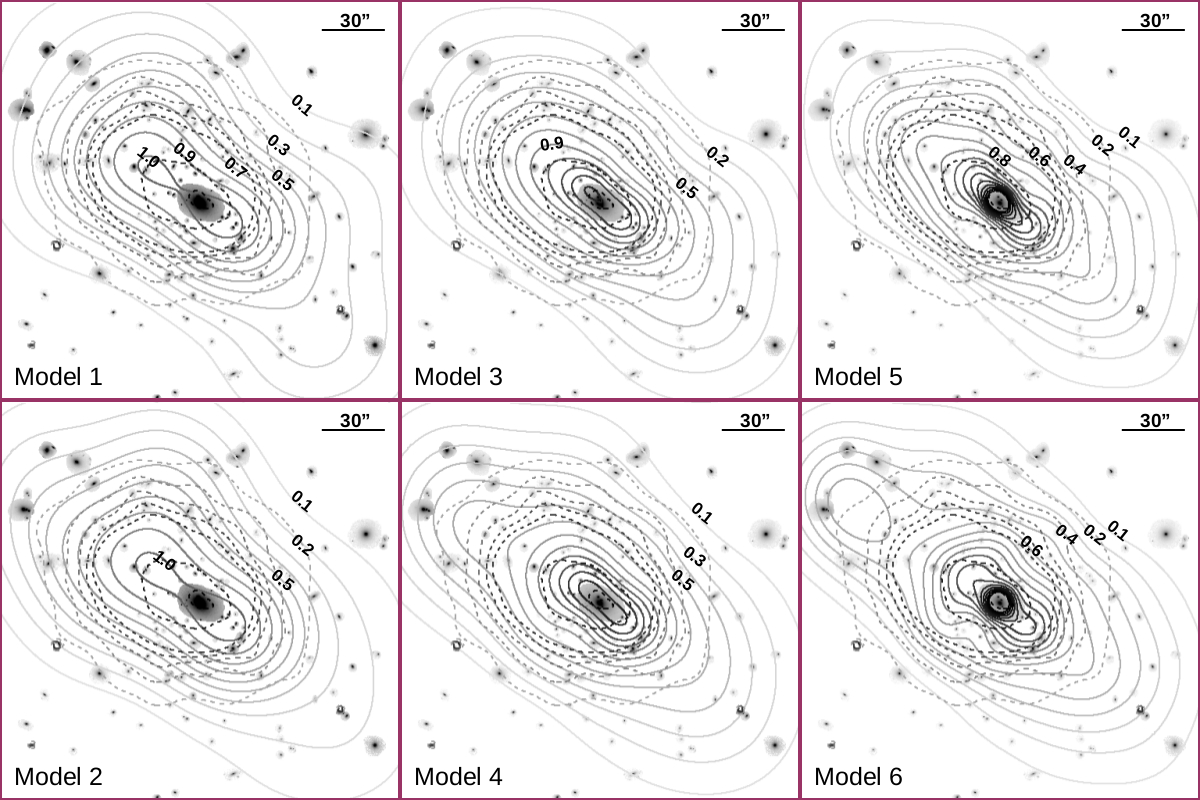}}  
   \caption{Mass contours of the soft component for the six solutions in units of the convergence ,$\kappa$, (for $z_s=3$, solid lines) compared with the X-ray contours from 
            Chandra (dashed lines) and the member galaxies (grey scale). The number in the contours indicate the value of $\kappa$.}
   \label{fig_Contours}  
\end{figure*}  

As shown in \cite{Diego2005,Diego2007}, the strong and weak lensing problem can be expressed as a system of linear
equations that can be represented in a compact form, 
\begin{equation}
\Theta = \Gamma X, 
\label{eq_lens_system} 
\end{equation} 
where the measured strong lensing observables (and weak lensing if available) are contained in the
array $\Theta$ of dimension $N_{\Theta }=2N_{SL}$, the
unknown surface mass density and source positions are in the array $X$
of dimension $N_X=N_c + N_g + 2N_s$ and the matrix $\Gamma$ is known
(for a given grid configuration and fiducial galaxy deflection field) 
and has dimension $N_{\Theta }\times N_X$.  $N_{SL}$ is the number
of strong lensing observables (each one contributing with two constraints,
$x$, and $y$), $N_c$ is the number of grid points (or cells) that we use to divide
the field of view. Each grid point contains a Gaussian function. The width of the Gaussians are chosen in such a way 
that two neighbouring grid points with the same amplitude produce a horizontal plateau in between the two 
overlapping Gaussians. In this work we consider different types of grid configurations. One of them is a regular grid 
with $N_c=16\times16=256$ grid points. In addition to the regular grid we consider also two multiresolution grids with 280 and 576 grid points with 
the resolution increasing gradually towards the BCG. The change in the grid configuration is one of the largest sources of variability on the reconstructed 
solutions. The different grid configurations cover the range of solutions where no prior information is given about the mass distribution (regular grid) and 
where a natural prior is given with a enhancement in the mass around the BCG. 
$N_g$ is the number of deflection fields (from cluster members) that we consider.  
In this work we set $N_g$ equal to 2. The first deflection field contains the BCG galaxy and the second deflection field contains the remaining 
galaxies from the cluster that are selected from the red-sequence (elliptical galaxies in the cluster)
Dividing the cluster galaxies in 2 layers allows us to 
independently fit the mass of the giant elliptical from the other galaxies. The particular configuration of the galaxies 
is shown in figure \ref{fig_Contours}. 
Finally, $N_s$ is the number of background sources (each contributes with two unknowns, 
$\beta_x$, and $\beta_y$) which in our particular case is $N_s=11$ when only the spectroscopic systems are used or  $N_s=35$ when 
all systems in Table~\ref{tab1} (up to system 45) are used in the reconstruction.
The solution is found after minimising a quadratic function that estimates the solution of the
system of equations (\ref{eq_lens_system}). For this minimisation we
use a quadratic algorithm which is optimised for solutions with the
constraint that the solution, $X$, must be positive. Since the vector $X$ contains the grid masses, 
the re-normalisation factors for the galaxy deflection field and the background source positions, and all these 
quantities are always positive (the zero of the source positions is defined in the bottom left corner of the 
field of view), imposing  $X>0$ helps in constraining the space of meaningful solutions. 
The condition $X>0$ also helps in regularising the solution as it avoids large negative and positive 
contiguous fluctuations. The quadratic algorithm convergence is fast (few minutes) on a desktop allowing 
for multiple solutions to be explored on a relatively sort time. Different solutions can be obtained after modifying 
the starting point in the optimization. A detailed discussion of the quadratic algorithm can be found 
in \cite{Diego2005}. A discussion about its convergence and performance can be found in \cite{Sendra2014}.

\section{Results}\label{sect_S7}
We apply the WSLAP+ algorithm to AS1063 following a typical strategy where a robust first version of the lens model based on the subset of systems with spectroscopic redshifts is built and used to identify new systems and constrain the redshift of systems with no spectroscopic redshift. We refer to this model as the {\it driver} model. 
Although photometric redshifts are available for some of these systems, in some cases photometric redshifts are significantly different from the true redshift. 
To avoid possible biases introduced by unreliable photometric redshifts, we rely instead on redshifts derived by the driver lens model.  These redshifts are more 
accurately predicted in systems where two images lie close to the critical curve. In this case, the critical curve constrains with accuracy the redshift 
of the system. In other cases, when the system is resolved, the distribution of morphological features, or knots, in the image plane can be used to constrain the 
redshift with great accuracy. An example is shown below for the new system 10 in section~\ref{Sect_Syst10}.

\begin{figure}  
 \centerline{ \includegraphics[width=10cm]{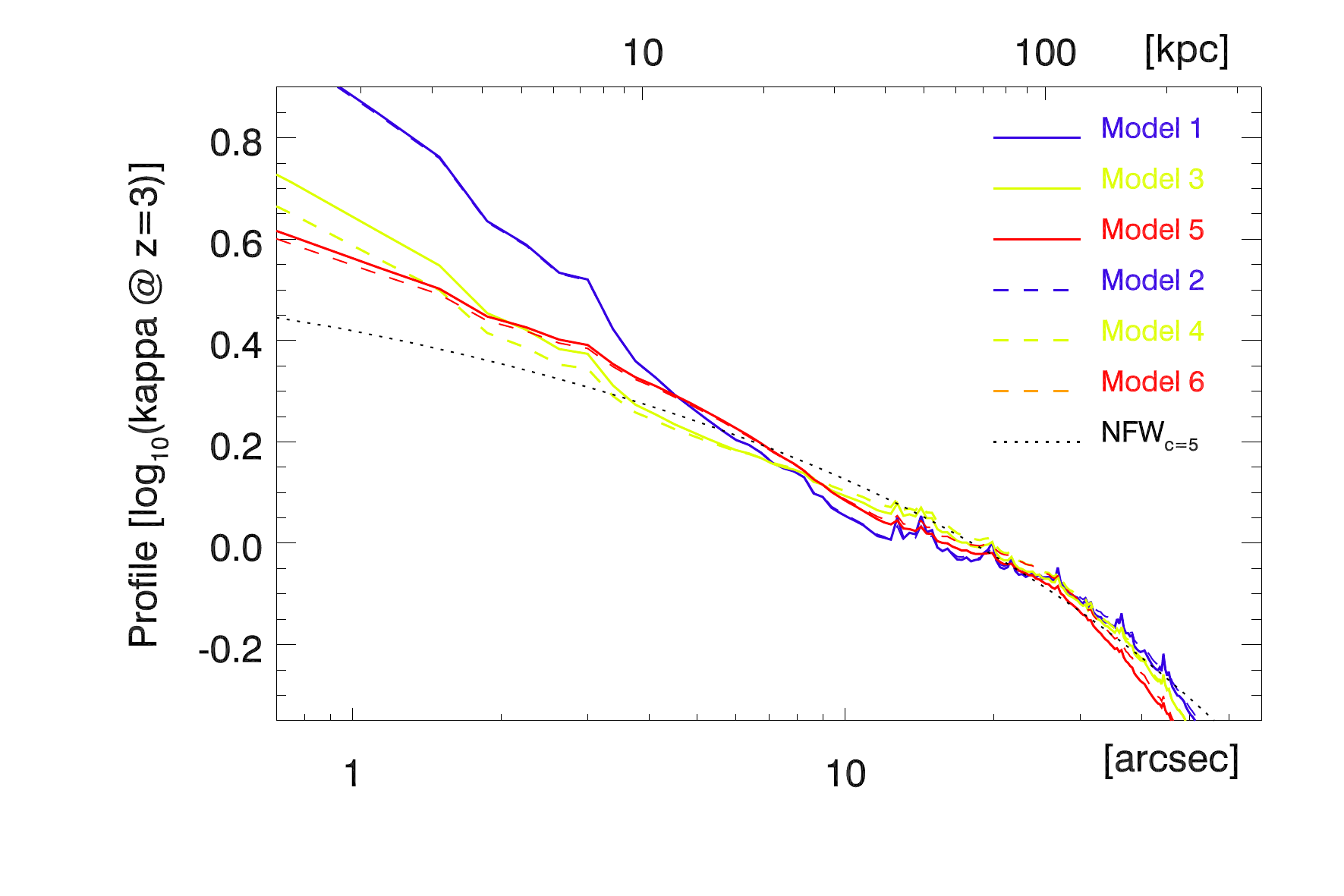}}  
   \caption{Mass profile for the different solutions compared with a NFW model. Models 1 and 2 are nearly indistinguishable in this plot. Models 3 and 4 are 
            also very similar but can be better distinguished  in the plot.} 
   \label{fig_Profiles}  
\end{figure}

\subsection{First guess and driver solution}\label{Sect_Driver}
An initial solution is derived using a regular grid for the soft component, the 11 spectroscopic systems shown in 
Fig.~\ref{fig_AS1063_Chandra}, and using the central knots as constraints.  This produces a preliminary solution.  
As discussed above and in more detail in \cite{Diego2016}, adding the spatial information of the systems that are resolved improves 
the solution and reduces its variability. AS1063 displays several well resolved systems (with spectroscopic redshift) which can be used to increase the number of constraints.  
In particular, we add spatial information for systems 1 (4 knots), 2 (3 knots), 5 (3 knots), 6 (6 knots), system 10 (2 knots), and system 19 (4 knots). 
Based on the preliminary solution we derive the size and orientation of the delensed images of these systems starting from the counterimage that is the least distorted 
in the image plane. The distribution of the knots 
in the delensed image with respect to the central knot is incorporated into the algorithm as additional constraints (leaving the position of the central knot 
as a free variable). With the addition of the new constraints we derive the {\it driver} solution that is used to identify new multiply lensed systems and to constrain 
the redshift of the new and previously known systems (with no spectroscopic redshift). 

\begin{figure*}    
 {\includegraphics[width=16.0cm]{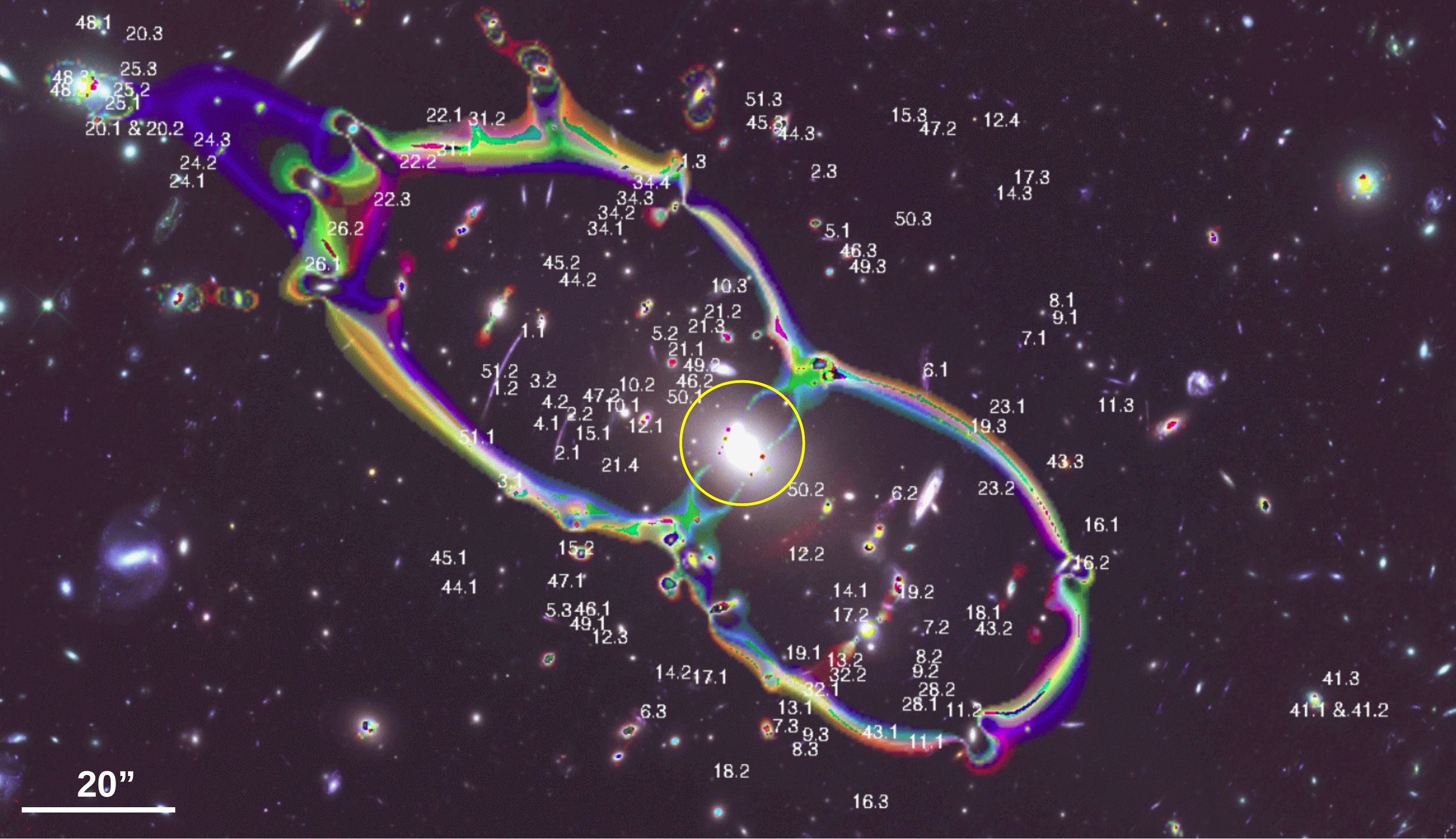}}
   \caption{Systems listed in Table~\ref{tab1} together with the critical curves from the six solutions. In red we show solutions 1 and 2, in green solutions 3 and 4 
            and in blue solutions 5 and 6. The yellow circle in the centre has a radius of $8''$ and marks the region that is poorly constrained by the data. }
   \label{fig_SystCritCurves}  
\end{figure*}

\subsection{New systems and the case of new system 10}\label{Sect_Syst10}
We use the driver model to i) identify new multiple lensed systems\footnote{All these systems are listed in Table~\ref{tab1}. 
Stamps of all entries in Table~\ref{tab1} can be found in http://www.ifca.unican.es/users/jdiego/AS1063} 
and ii) to predict the redshift of the new systems and of previously known systems with no spectroscopic 
redshift. The list of new systems and redshifts is given in Table~\ref{tab1} in the appendix and are marked with an asterisk. 
Although the redshifts listed in Table~\ref{tab1} include the five new spectroscopic redshifts for systems 7, 10, 13, 14, and 19 compiled in  
\cite{Caminha2015a} (see references in Table~\ref{tab1} for proper credit of the redshift measurements), 
it is interesting to compare the values of the new spectroscopic redshifts with the blind redshift estimates provided by the lens model before the paper 
by \cite{Caminha2015a} was published. That is, we predict the new spectroscopic redshifts using a model that is constrained by the 6 systems with previously 
known spectroscopy (systems 1, 2, 5, 6, 11 and 12).  
For the four previously known systems with new spectroscopic redshifts (7, 13, 14, and 19) this lens model (copnstrained by systems  1, 2, 5, 6, 11 and 12) 
predicted $z_{model}=1.9, 3.5, 3.2$ and 1.05 respectively whereas the measured spectroscopic redshift is $z_{spect}=1.837, 4.113, 3.118$ and 1.035 respectively. 
In addition, our new system 10 (which is not recognised as a lensed system by any previous work) had a predicted $z_{model}=0.78$ while in \cite{Caminha2015a} the 
brightest counterimage of this system has a  spectroscopic redshift $z_{spect}=0.73$ in good agreement with our prediction.
System 10 is a good illustration of the power of the lens model to uncover new systems and to correctly predict their redshifts. 
Despite being clearly detected in previous images from the CLASH program, it was overlooked as a multiply lensed system 
by several authors, probably due to the fact that only a portion of a background galaxy is multiply lensed. 
A slightly different version of the driver model (similar to the driver model described in section \ref{Sect_Driver} but excluding system 10 from the constraints) 
correctly predicts the redshift and the morphology of the lensed image as shown in Fig.~\ref{fig_Syst10}. 
The sensitivity to the redshift is impressive and is comparable to the precision attained by photometric redshifts. A photometric redshift estimate (made a posteriori) of the relatively of the galaxy in the north-west part of Fig.~\ref{fig_Syst10}  results in $z_{phot}=0.697^{+0.063}_{0.037}$, in good agreement (at 1 $\sigma$) 
with the redshift inferred from the lens model and the spectroscopic redshift. 
Our lens model predcics a redshift ($z_{geom}=0.78$) that is biased high with respect to the spectroscopic redshit ($z_{spect}=0.73$). 
The comparison (discussed at the beginning of this section) between the new spectroscopic redshifts in \cite{Caminha2015a} and the blind estimates based on the lens model that relied on the original 6 systems with spectroscopic redshift shows a similar systematic bias that pushes the geometric redshifts towards slightly larger values. 
This systematic bias is also identified in other systems with known spectroscopic redshift like in system 2 where we find that the model prefers a redshift 
of z=1.285 as opposed to the observed z=1.261 in order to accurately reproduce the observed arc (counterimages 2.1 and 2.2, see Fig.~\ref{fig_Relensed}). 
For this system, when we assume z=1.261, the predicted counterimages 2.1 and 2.2 based on the observed 2.3 merge into a vanishing arc similar to the situation when z=0.77 
is assumed for system 10 (see bottom-left panel in Fig,~\ref{fig_Syst10}). 
Including system 10 in the driver model (with the spectroscopic redshift) alleviates the tension between the model and observed redshift 
of systems 2 and 10 but does not eliminate the bias. System 2 still prefers z=1.28 (as opposed to the measured z=1.261) and system 10 prefers z=0.75 
(as opposed to the measured z=0.73). 
Interestingly, the bias seems to be more obvious around the region where there is only systems with relatively low spectroscopic redshifts (see Fig.~\ref{fig_AS1063_Chandra}). 
This possible systematic bias will be investigated in the future in this and other clusters but it was not observed in our previous works suggesting that it may be intrinsic to this 
cluster and perhaps linked with the lack of central constraints or the lack of high redshift constraints in this part of the lens. 
  
Using the driver model obtained with the 11 spectroscopic redshifts we unveil new systems together with their geometric redshifts predicted by this model (together with the redshifts of previously known systems that did not have spectroscopic redshift). The new systems are discovered thanks to the increased depth of the data but also thanks to the success of the driver model at correctly predicting the position and redshifts of the different counterimages. Several of the new systems may not be multiply lensed systems but rather very elongated arclets. In this case we quote several positions along this arc that are later used as constraints in WSLAP+. Four such systems (22, 24, 26 and 34 in Table~\ref{tab1}) are included in our set of constraints.   
With the driver model we identify a total of 35 systems listed in Table~\ref{tab1}. The identification is made after matching the positions, 
colors and morphology of the observed and predicted (by the driver model) images. 
Using an updated model derived with these 35 systems, we later discover 7 additional system candidates also listed in Table~\ref{tab1}. 
Examples of the model predictions for some of these (spatially resolved) systems are given in Fig~\ref{fig_Relensed}. 
Systems that are not spatially resolved do not provide useful information when comparing the model and the observed images 
(other than comparing positions). 

The agreement between the predicted and observed images is remarkable except for image 19.2 (and 2.1, 2.2 as explained above). 
For this particular image (19.2), the presence of a nearby spiral galaxy (seen in the north part of the stamp) introduces a small scale distortion that is not properly 
captured by our model that includes only elliptical member galaxies. 
The spiral galaxy was later introduced although not as an extra free-parameter, but locking its luminosity-to-mass relation to that of the elliptical galaxies. 
The image shown in Fig~\ref{fig_Relensed} already includes this spiral galaxy in the model and helps in better reproduce this image, but there 
is still a residual error, probably linked to the constraint that the mean light-to-mass ratio of the model is locked to that of the elliptical galaxies used in this model. 

Using the systems and redshifts compiled in Table~\ref{tab1}, we describe in the next section the set of lens models that are derived from these constraints. 

\begin{table}
  \begin{minipage}{80mm}                                               
    \caption{Integrated total mass as a function of radius. The mass is given in units of $10^{14} M_{\odot}$.
             The mean and dispersion is computed from the six models described in section~\ref{sect_S6}}
 \label{tab0}
 \begin{tabular}{| c c c || c c c |}   
 \hline
  R(kpc)  &    M($<R$)  &  $\sigma$  &  R(kpc)  &    M($<R$)  &  $\sigma$  \\
 \hline
   0.92   &  0.0007     &  0.0003    &  141.1   &    1.343     &   0.016   \\
   2.76   &  0.0033     &  0.0015    &  163.2   &    1.673     &   0.022   \\
   6.45   &  0.0134     &  0.0053    &  185.4   &    2.012     &   0.033   \\
  10.14   &  0.0282     &  0.0096    &  207.5   &    2.346     &   0.049   \\
  15.68   &  0.052      &  0.014     &  229.7   &    2.677     &   0.069   \\
  19.37   &  0.072      &  0.016     &  251.8   &    2.990     &   0.090   \\
  24.90   &  0.101      &  0.017     &  273.9   &    3.291     &   0.111   \\
  30.44   &  0.133      &  0.018     &  296.1   &    3.578     &   0.133   \\
  37.82   &  0.183      &  0.018     &  318.2   &    3.849     &   0.155   \\
  45.20   &  0.235      &  0.017     &  340.4   &    4.105     &   0.178   \\
  54.42   &  0.307      &  0.014     &  362.5   &    4.343     &   0.200   \\
  63.65   &  0.386      &  0.011     &  384.6   &    4.568     &   0.223   \\
  74.72   &  0.495      &  0.007     &  406.8   &    4.771     &   0.245   \\
  96.86   &  0.740      &  0.006     &  428.9   &    4.962     &   0.270   \\
  119.0   &  1.025      &  0.011     &  451.1   &    5.137     &   0.295   \\
 \hline
 \end{tabular}
    \end{minipage}
\end{table}

\subsection{Models}\label{sect_S6}

\begin{figure*}    %
 {\includegraphics[width=16.0cm]{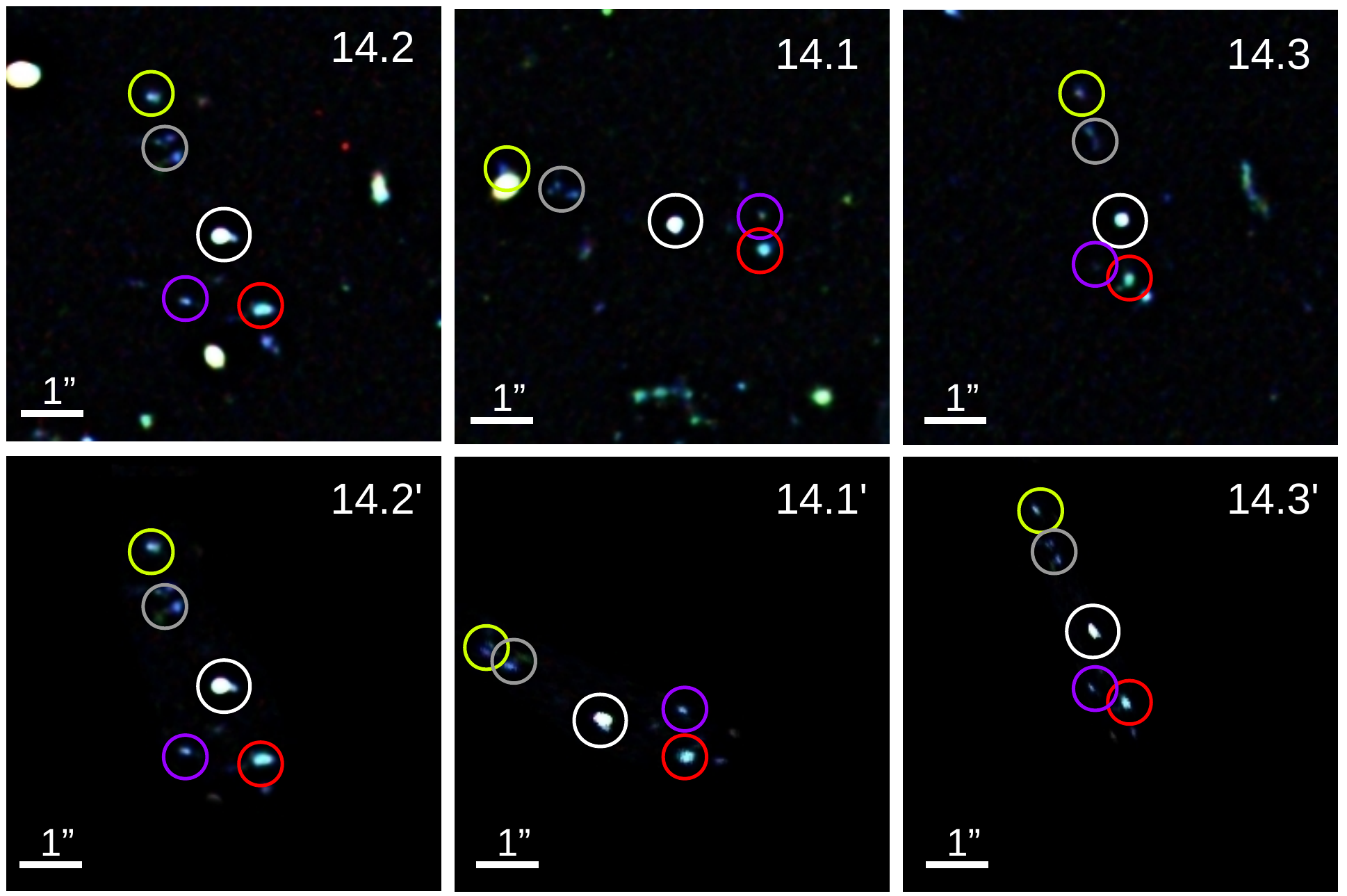}}
   \caption{Extra galaxies around the galaxy in system 14 (z=3.118). The data is shown in the top row and the model in the bottom row. 
            Each galaxy is marked with a different colour to facilitate their 
            identification. The original system 14 is marked with a white circle. The model images are constructed based on image 14.2 and 
            predicting the image in the image plane assuming all the galaxies are at the same redshift. Each stamp is centered in the exact same 
            position. The good agreement between the data and the model indirectly confirms that these additional galaxies are at the same redshift 
            as system 14. The delensed image spans over $\approx 3''$ which corresponds to a physical size of $\approx 23$ kpc at the distance 
            of system 14. } 
   \label{fig_Syst14}  
\end{figure*}  

To account for uncertainties and variability in the solutions, we explore a range of cases (or models) 
where we change the grid configuration, which we find accounts for the largest source of variability in the derived solutions, and we also vary the set of systems used to perform the reconstruction. In particular we consider six types of models (or cases) described briefly below.\\
 
$\bullet$ Case 1. We use a standard grid of $16\times16=256$ cells in our field of view. We use only the systems with spectroscopic redshifts to perform the 
                  lens reconstruction. The optimization algorithm is iterated 50000 times. This case corresponds to the driver model referred to above.\\ 
$\bullet$ Case 2. Like Case 1 but we use all the systems listed in Table~\ref{tab1} in the appendix (except those marked with **). The optimization algorithm is iterated 150000 times.  
                  Note that systems marked with ** in  Table~\ref{tab1} are highly consistent with the model by construction and hence have little extra constraining power.\\
$\bullet$ Case 3. Like Case 1 but instead of a uniform regular grid we use a multiresolution grid with 280 cells.\\
$\bullet$ Case 4. Like Case 3 but we use all the systems listed in Table~\ref{tab1} in the appendix (except those marked with **). The optimization algorithm is iterated 150000 times.  \\
$\bullet$ Case 5. Like Case 3 but instead of a multiresolution grid with 280 cells we increase the resolution and use a grid with 576 cells. \\
$\bullet$ Case 6. Like Case 5 but we use all the systems listed in Table~\ref{tab1} in the appendix (except those marked with **). The optimization algorithm is iterated 150000 times.  \\

In all cases, we assume two deflection fields for the galaxies as described in the previous section. 
The BCG is treated as an independent deflection field and its mass is re-scaled by the algorithm in the minimization process. 
For the remaining cluster members, their masses are equally re-scaled by the same factor (but different from the factor for the BCG). 

For each one of the six cases discussed in section \ref{sect_S6} we derive a solution (mass distribution, position of background sources, 
deflection field at a fiducial redshift $z_f=3$, magnification maps and critical curves). 
The minimization is stopped once the solution has converged to a stable point (after 50000 or 150000 iterations). 

\subsection{Mass profile and mass distribution}
The integrated mass as a function of radius is given in Table~\ref{tab0}. 
Our integrated mass is in good agreement with other estimates given in \cite{Richard2014} and \cite{Zitrin2015} from very different parametric based models. 
Comparing our results with the values found in the literature within 250 kpc, we obtain $M(<250 kpc) = (2.97 \pm 0.09)\times10^{14}$ whereas 
\cite{Monna2014} and \cite{Johnson2014} find slightly lower masses of $(2.67 \pm 0.08)\times10^{14}$ and 
$(2.68^{0.03}_{-0.05})\times10^{14}$, respectively, and \cite{Caminha2015a} finds a closer value to ours of $(2.9 \pm 0.02)\times10^{14}$. 
The lower masses found by \cite{Monna2014} and \cite{Johnson2014} may be due in part to a lack of spectroscopic information for some systems 
leading to an underestimation of the total mass. 
A comparison of the convergence profiles (computed as the surface mass density divided by the critical surface mass density at z=3) 
for the six models is shown in Fig. \ref{fig_Profiles}. 
The agreement between the profiles is very good in the range 20-200 kpc, which is the range covered by 
the lensing constraints.  In this range, we find that a NFW profile with a concentration parameter $C \approx 5$ and  total mass of $1.93 \times 10^{15} M_{\odot}$ 
(within a radius r=1.5 Mpc) produces a good match to the observed projected profile, consistent with the 
hypothesis that this cluster may be relatively relaxed. According to results from simulations \citep{Meneghetti2014} and recent observations of clusters  
\citep{Merten2014}, massive galaxy clusters are well reproduced by NFW profiles with relatively low values for the concentration parameter of $C \approx 3-4$, 
although somewhat larger values are derived for well defined relaxed clusters from the CLASH program \citep{Umetsu2014,Zitrin2015}.

The 2-dimensional distribution of the soft component (grid) for the mass is shown in Fig.~\ref{fig_Contours} together with the position (and shape) 
of the input galaxies as well as the X-ray emission from  {\it Chandra}.
The peak of the soft component aligns well with the position of the BCG.  
In the cases of the multi-resolution grid, this alignment may be a consequence of the  prior introduced by 
the grid rather than a well constrained result. The total mass profile (soft component plus compact component) 
in the centre is noticeably steeper for the solutions with the regular grid as shown in Fig~\ref{fig_Profiles}.
The mass distribution is elongated in the diagonal direction towards what seems to be a secondary clump in the north-east. This clump is more evident in the case of 
model 6 in Fig.~\ref{fig_Contours}. 
Some prominent members of the cluster are found also near the location of this clump. The elongation increases towards the centre of the cluster, in agreement with expectations 
from N-body simulations \citep{Allgood2006}. In particular, we find axis ratios $a/b \approx 1.9$ in the convergence isocontour $\kappa = 0.2$ that increases up to 
 $a/b \approx 2.3$ for  $\kappa = 1.0$ (from models 4 and 6) where $a$ and $b$ are the largest and smallest axis respectively.

\subsection{Additional systems}
Using the refined lens models derived above and the filtered version of the color image we search for additional systems. We find  7 new systems listed at the 
end of Table~\ref{tab1} (below system 45). In addition to these 7 systems, we identified one extra system (system 52) after finding the counterimage for one 
background galaxy with new spectroscopic redshift in \cite{Caminha2015a}.  
Stamps of the new systems are also provided in the webpage\footnote{http://www.ifca.unican.es/users/jdiego/AS1063} with supporting material. 
Although these 7 new additional systems are not used to further constrain the solution, adding them should have a minimal impact  
since these systems are highly consistent with the model derived to identify them.
Among the new systems, one of them is probably associated with system 14.  
System 14 is surrounded by a number of small galaxies that seem physically associated with the central knot (system 17 also lies very close in the source plane with a redshift consistent with that of system 14). The best redshift predicted by the lens models for these small galaxies is the same as that of system 14. 
Assuming that all these galaxies are at the same redshift ($z=3.118$), the predicted lensed images in the image plane agree remarkably well with the observed distribution of galaxies (see Fig.\ref{fig_Syst14}). 
If these galaxies are in fact physically associated with system 14, they would form a structure of $\approx 23$ kpc at redshift $z=3.118$ aligned almost 
perfectly along a straight line. The size of the galaxies around system 14 is about 1 kpc, which agrees well with the typical half mass radius of star forming galaxies at z=3 (see for instance \citep{Oser2012}). This is well below the observed 23 kpc for the entire structure, implying that the observed structure at z=3.118 may then be a collection of different star forming galaxies at this redshift. 
\begin{figure*}    
 {\includegraphics[width=16.0cm]{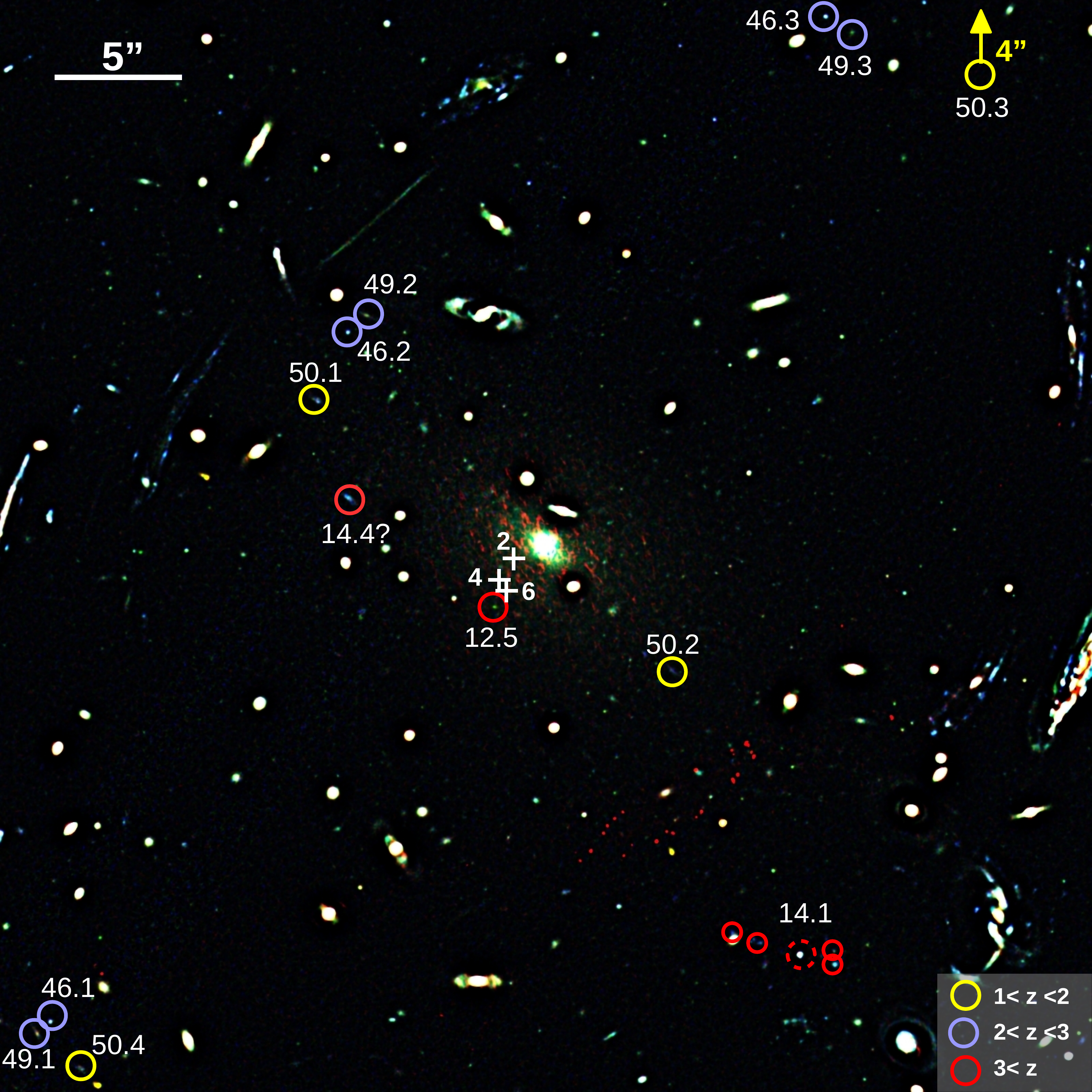}}
   \caption{Central $0.8'\times0.8'$ region. Additional candidate systems are marked with circles. The predicted position of a central 5th predicted image 
            12.5 is marked with a white cross for models 2,4 and 6. 
            A reddish candidate is found within a fraction of an arcsec of that position (red circle near the BCG). 
            The red circles next to image 14.1 mark the positions of galaxies that seem to be at the exact same redshift of system 14 and 
            could be sub-structures linked to a larger structure at that redshift. This color image is produced after filtering out the large scale diffuse light from the cluster and 
            individual galaxies. A smoothing has been applied also to boost the signal-to-noise.} 
   \label{fig_Centre}  
\end{figure*}  
An interesting possibility, and complementary to the above, is that we may be witnessing the star formation in these galaxies that is being (or has been) triggered by a jet from an AGN in the central, brightest knot in system 14 \citep{Gaibler2012}. The alignment and symmetry of the new galaxies around system 14 supports this hypothesis. Similar examples are known of such linear alignment of galaxies spanning nearly 20 kpc at redshift of z=3 \citep{Rauch2013}. 
Also in the local universe there are examples of star forming regions that are being triggered by jets and be found as far as 20 kpc \citep{vanBreugel1985} or even 70 kpc \citep{Salome2015} from the AGN. 
None of the central knots of system 14 (in the 3 multiple images) show any excess X-rays in Chandra data despite being magnified by factors between 
2.5 and 6.5 implying in this scenario that the AGN is weak or in a quiescent state. However, we should recall again that the data used in this work is relatively shallow (26.7 ks). System 14, together with the galaxies aligned with it resembles what are known as ''chain galaxies''. In fact,  chain galaxies are a well-established phenomenon, going back to \cite{Cowie1995}, and defined to be linearly organized chains of giant HII regions seen in galaxies in  the faintest HST WFPC2 images.
A recent example, discovered via  a strongly lensed cluster, is a thirty kiloparsec chain of star-forming "beads on a string"  \citep{Tremplay2014} although this is attributed to a merger between two early-type galaxies in the cluster core.

Such an alignment as we find in  System 14 may be due to  the rare case of  an orientation effect associated with  a  star forming disk galaxy being viewed edge-on, combined with surface brightness dimming at high redshift, as found in deep HST images by \cite{Elmegreen2004} and \citep{Elmegreen2005}. 
However there is an alternative and possibly more compelling interpretation.
AGN-driven nuclear ionized outflows are ubiquitous in the most massive star-forming galaxies at high redshift \citep{Genzel2014}. This high duty cycle phenomenon is a prime candidate for triggering of star formation due to a  narrow ionization cone or jet as observed by \cite{Cresci2015}. 
We  cite two other examples suggestive of triggering by positive feedback. One is the case of the possible one-sided jet AGN  zC400569 plus aligned massive clumps at $z \sim 2$ mapped in \cite{Forster2014}. A second is a series of massive CO clumps aligned along a quasar jet at z=4.3, providing triggering of  molecular gas (on scales up to $\approx$ 15 kpc), the essential prerequisite for star formation, in \cite{Klamer2004}. Such offsets or alignments are  inferred to be a common phenomenon in CO-detected high redshift radio sources \citep{Emonts2014}. Finally, we note that theoretical  simulations  of  positive feedback by nuclear jets \citep{Gaibler2012} and winds \citep{Wagner2013} support this general picture. 
At the time of finalizing this paper, we noticed the intersting work of \cite{Caminha2015b} where system 14 is discussed in more detail and throw more light into this peculiar high-redshift object. They show how the central blob in system 14 is surrounded by a $Ly-\alpha$ nebula of 33 kpc in size. They conclude that the  $Ly-\alpha$ nebula is probabbly powered by embedded star formation. Finally, they confirm that the redshift of the galaxies aligned with the central blob is the same in agreement with our original assumption.

\begin{figure*}    %
 {\includegraphics[width=16cm]{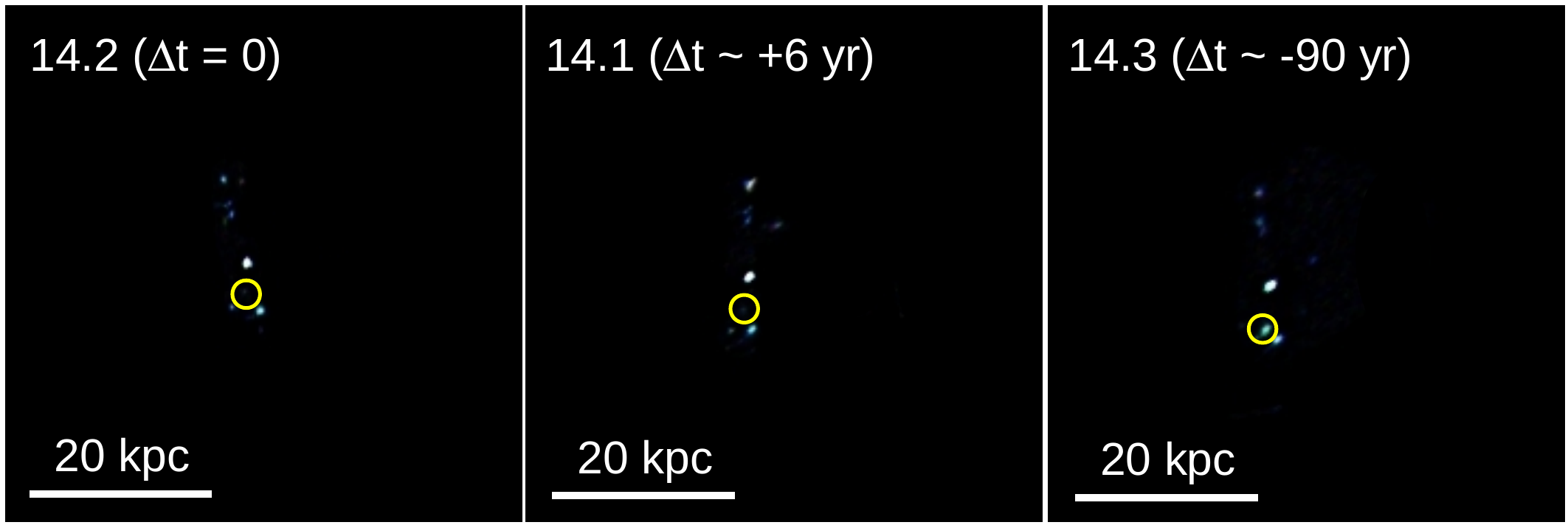}}
   \caption{Predicted images of system 14 and surrounding galaxies in the source plane assuming all galaxies are at the same redshift (z=3.118). 
            The galaxies around system 14 align well in the same direction in the source plane. In the stamp corresponding to 14.3, the southernmost galaxy 
            is significantly brighter. This could be explained if significant variability in this source (linked with the alleged jet) takes place on scales 
            of less than 90 years given the expected lensing time delay.} 
   \label{fig_Syst14_source}  
\end{figure*}

\section{Discussion}\label{Sect_Discus}
The mass profiles shown in Fig.~\ref{fig_Profiles} are unconstrained within a radius of 20 kpc due to the lack of any clear central lensed 
image in this area. The mass profile we derive and shown 
in  Fig.~\ref{fig_Profiles} for cases 1, and 2 would imply a surprisingly high concentration parameter that may be difficult to reconcile with the predictions  emerging from 
N-body simulations as well as other observational constraints \citep{Meneghetti2014,Merten2014,Umetsu2014,Zitrin2015}. 
We can, however, may turn this around and explore whether the absence of central lensed images can help limit the gradient of the inner mass profile  near the BCG. The possible exception of image 12.5 is discussed below separately. We find that profiles that are steeper in the centre, like models in cases 1 and 2, predict a counter image in the central region for system 5 with a magnification of $\mu \approx 0.75$ that should be prominent in the data but is  not currently observed. 
At $\approx 1''$ from the predicted image location there is an arc (image 50.2 in Fig.~\ref{fig_Centre}) with similar orientation to the predicted one 
but with very different colors. Similarly for system 6, a relatively bright central image ($\mu \approx 0.6$) is predicted but not found. These central images are not predicted by the shallower models derived with the multiresolution grid (cases 3,4,5,6). 
For system 14, models 1 and 2 predict also a relatively bright ($\mu \approx 0.8$) central counterimage. In this case a bright candidate (marked with 14.4? in Fig.~\ref{fig_Centre}) 
is found at less than $1''$ distance from the predicted position but again with colours that differ from the observed counterimages. Unless 14.4 is confirmed as a genuine counterimage, the lack of these
central images seems to favour a shallower profile like those of our models 3, 4, 5 and 6. As an additional test, if we adopt as reasonable the counter image 12.5 and use it as an extra constraint, the solution also favours the shallower profiles as they better predict the position of this counterimage as shown in  Fig.~\ref{fig_Centre}. Image 12.5 will be useful in the future to better constrain the central region. We should note that other models, like the one found by \cite{Caminha2015a}, also predict a counterimage very close to the candidate 12.5. A spectroscopic confirmation of this candidate will be available soon (Karman et al. in prep.).

As shown in Fig.~\ref{fig_Contours}, we find that the dark matter follows an elongated distribution in the diagonal direction. The same elongation and orientation is found in weak lensing analysis of the same cluster on scales up to 30 arcminutes \citep{Gruen2013}. 
The elongation is in the direction of a second clump in the north-east corner of the field of view. This clump is most evident in model 6. The smaller number of constraints
around this clump does not allow for a detailed mapping of the clump but its existence is supported by the presence of several prominent galaxies in 
its vicinity. The fact that the distribution of the X-rays seems to be elongated in the same direction and  the centroid of the X-ray contours 
is located between the main cluster and the secondary clump  also supports the hypothesis of a perturbing second clump. On the other hand, 
no excess X-rays are found at the position of the second clump, although it should be kept in mind that the data used in this work is relatively shallow.
The degree of elongation seems to increase towards the central region of the cluster. This is expected in clusters modelled in the standard framework \citep{Allgood2006}, where such obvious large elongations are  expected for the most massive clusters.

 The possibility that the extra galaxies found around system 14 are seen due to jet-induced star formation, could be directly examined with deep radio observations of the 
radio emission associated with the alleged jet. Unfortunately, to the best of our knowledge, no deep radio data is available for this cluster. 
When comparing the predicted distribution of the galaxies around system 14 in the source plane we find that around 14.3, one of the galaxies is significantly 
brighter than the other two (marked with a yellow circle in Fig.~\ref{fig_Syst14_source}).
To explore the jet-induced star formation scenario, we compute the time delays between 
the different counterimages of system 14. The lens model predicts that the light from 14.3 arrives approximately 90 years before the light in 14.2 and approximately 96 years 
before the light observed in 14.1 (the uncertainty in the time delays is about 2 years based on the dispersion of the six lens models). 
Consequently, if the galaxy marked with a yellow circle in Fig.~\ref{fig_Syst14_source} is indeed at the same redshift as system 14, the jet-induced star formation in the galaxy in 14.3 is unlikely to be responsible for its relative brightness. We can not, however, rule out sort-lived events, like supernova for which the formation of their progenitor stars may have been triggered by the jet. As in the case of the recent supernova Refsdal \citep{Diego2016,Kelly2016}, this scenario can be tested in the future although we will have to wait for $\approx 90$ years. Other less exotic explanations, like microlensing, could be worth exploring.

\section{Conclusions}\label{sect_S8}

Using the latest optical images from the HFF programme (in the F435W, F606W and F814W filters) 
we have unveiled new multiply lensed systems and constrained their redshifts, bringing the total number of 
(candidate) lensed background galaxies to more than 40 and the number of multiply lensed images to more than 100. 
We derive six mass models for the cluster, spanning a range of the most important variables, and compare the resulting mass profiles, projected mass distributions and critical curves. The models agree well with each other. We find the largest differences between models in the central region. This is consistent with the lack of centrally lensed images. By requiring 
that some systems do not produce a central counterimage, we infer that the best models are those with shallower slopes in the central region. In particular, 
an NFW model with concentration parameter $C \approx 5$ agrees well with the observations.  The mass distribution is elongated in the diagonal direction, with one of the models 
showing a clump in the north-east at approximately 470 kpc from the BCG. Several prominent galaxies are found near this clump suggesting that it may be a real dark matter feature in 
the lens. X-ray (shallow) data from Chandra does not show any excess of X-rays at the position of the clump. The elongation, however, is supported by independent weak 
lensing results and is consistent also with previous studies based on strong lensing. The elongation increases towards the centre of the cluster in agreement 
with expectations from N-body simulations for such a massive cluster. Among the newly discovered multiple images, we find several distant galaxies at the same redshift of system 14 (z=3.118) 
that form a linear structure spanning 23 kpc with system 14 at its centre. We discuss the possibility that this system is an example of jet-induced star formation at redshift $z \approx 3$, 
an scenario that can be tested with radio obsrvations. 

\section{Acknowledgments}  
This work is based on observations made with the NASA/ESA {\it Hubble Space Telescope} and operated by the Association of Universities for Research in Astronomy, Inc. 
under NASA contract NAS 5-2655. 
Part of the data for this study is retrieved from the Mikulski Archive for Space Telescope (MAST).
The authors would like to thank the HFF team for making this spectacular data set promptly available to the community.
The scientific results reported in this article are based in part on data obtained from the Chandra Data 
Archive \footnote{ivo://ADS/Sa.CXO\#obs/4966}
We would like to thank Harald Ebeling for making  the code {\small ASMOOTH} \citep{Ebeling2006} available. 
T. J. Broadhurst gratefully acknowledges the Visiting Research Professor Scheme at the University of Hong Kong.
J. Lim acknowledges a grant from the Seed Funding for Basic Research of the University of Hong Kong (Project No. 201411159166) for this work.
J.M.D acknowledges support of the consolider project CSD2010-00064 and AYA2012-39475-C02-01 funded by the Ministerio de Economia y Competitividad. 
The work of JS was supported at IAP by  ERC Project No. 267117 (DARK) hosted by the Pierre and Marie Curie University- Paris 6, and at JHU by NSF grant OIA-1124403. JMD acknowledges the hospitality of the Physics Department at University of Pennsylvania for hosting him during the preparation of this work. 
The authors thank G.B. Caminha, W. Karman and K.B. Schmidt for useful comments and suggestions.  
  
\label{lastpage}
\bibliographystyle{mn2e}
\bibliography{MyBiblio} 

\begin{thebibliography}{54}
\expandafter\ifx\csname natexlab\endcsname\relax\def\natexlab#1{#1}\fi

\bibitem[{{Allgood} {et~al}\mbox{.}(2006){Allgood}, {Flores}, {Primack},
  {Kravtsov}, {Wechsler}, {Faltenbacher}, \& {Bullock}}]{Allgood2006}
{Allgood} B., {Flores} R.~A., {Primack} J.~R., {Kravtsov} A.~V., {Wechsler}
  R.~H., {Faltenbacher} A., {Bullock} J.~S., 2006, \mnras, 367, 1781

\bibitem[{{Balestra} {et~al}\mbox{.}(2013){Balestra}, {Vanzella}, {Rosati},
  {Monna}, {Grillo}, {Nonino}, {Mercurio}, {Biviano}, {Bradley}, {Coe},
  {Fritz}, {Postman}, {Seitz}, {Scodeggio}, {Tozzi}, {Zheng}, {Ziegler},
  {Zitrin}, {Annunziatella}, {Bartelmann}, {Benitez}, {Broadhurst}, {Bouwens},
  {Czoske}, {Donahue}, {Ford}, {Girardi}, {Infante}, {Jouvel}, {Kelson},
  {Koekemoer}, {Kuchner}, {Lemze}, {Lombardi}, {Maier}, {Medezinski},
  {Melchior}, {Meneghetti}, {Merten}, {Molino}, {Moustakas}, {Presotto},
  {Smit}, \& {Umetsu}}]{Balestra2013}
{Balestra} I. {et~al.}, 2013, \aap, 559, L9

\bibitem[{{Boone} {et~al}\mbox{.}(2013){Boone}, {Cl{\'e}ment}, {Richard},
  {Schaerer}, {Lutz}, {Wei{\ss}}, {Zemcov}, {Egami}, {Rawle}, {Walth}, {Kneib},
  {Combes}, {Smail}, {Swinbank}, {Altieri}, {Blain}, {Chapman},
  {Dessauges-Zavadsky}, {Ivison}, {Knudsen}, {Omont}, {Pell{\'o}},
  {P{\'e}rez-Gonz{\'a}lez}, {Valtchanov}, {van der Werf}, \&
  {Zamojski}}]{Boone2013}
{Boone} F. {et~al.}, 2013, \aap, 559, L1

\bibitem[{{Bozek} {et~al}\mbox{.}(2015){Bozek}, {Marsh}, {Silk}, \&
  {Wyse}}]{Bozek2015}
{Bozek} B., {Marsh} D.~J.~E., {Silk} J., {Wyse} R.~F.~G., 2015, \mnras, 450,
  209

\bibitem[{{Bradley} {et~al}\mbox{.}(2014){Bradley}, {Zitrin}, {Coe}, {Bouwens},
  {Postman}, {Balestra}, {Grillo}, {Monna}, {Rosati}, {Seitz}, {Host}, {Lemze},
  {Moustakas}, {Moustakas}, {Shu}, {Zheng}, {Broadhurst}, {Carrasco}, {Jouvel},
  {Koekemoer}, {Medezinski}, {Meneghetti}, {Nonino}, {Smit}, {Umetsu},
  {Bartelmann}, {Ben{\'{\i}}tez}, {Donahue}, {Ford}, {Infante}, {Jimenez-Teja},
  {Kelson}, {Lahav}, {Maoz}, {Melchior}, {Merten}, \& {Molino}}]{Bradley2014}
{Bradley} L.~D. {et~al.}, 2014, \apj, 792, 76

\bibitem[{{Caminha} {et~al}\mbox{.}(2015{\natexlab{a}}){Caminha}, {Grillo},
  {Rosati}, {Balestra}, {Karman}, {Lombardi}, {Mercurio}, {Nonino}, {Tozzi},
  {Zitrin}, {Biviano}, {Girardi}, {Koekemoer}, {Melchior}, {Meneghetti},
  {Munari}, {Suyu}, {Umetsu}, {Annunziatella}, {Borgani}, {Broadhurst},
  {Caputi}, {Coe}, {Delgado-Correal}, {Ettori}, {Fritz}, {Frye}, {Gobat},
  {Maier}, {Monna}, {Postman}, {Sartoris}, {Seitz}, {Vanzella}, \&
  {Ziegler}}]{Caminha2015a}
{Caminha} G.~B. {et~al.}, 2015{\natexlab{a}}, ArXiv e-prints

\bibitem[{{Caminha} {et~al}\mbox{.}(2015{\natexlab{b}}){Caminha}, {Karman},
  {Rosati}, {Caputi}, {Arrigoni Battaia}, {Balestra}, {Grillo}, {Mercurio},
  {Nonino}, \& {Vanzella}}]{Caminha2015b}
{Caminha} G.~B. {et~al.}, 2015{\natexlab{b}}, ArXiv e-prints

\bibitem[{{Coe}, {Bradley} \& {Zitrin}(2015){Coe}, {Bradley}, \&
  {Zitrin}}]{Coe2015}
{Coe} D., {Bradley} L., {Zitrin} A., 2015, \apj, 800, 84

\bibitem[{{Coe} {et~al}\mbox{.}(2013){Coe}, {Zitrin}, {Carrasco}, {Shu},
  {Zheng}, {Postman}, {Bradley}, {Koekemoer}, {Bouwens}, {Broadhurst}, {Monna},
  {Host}, {Moustakas}, {Ford}, {Moustakas}, {van der Wel}, {Donahue}, {Rodney},
  {Ben{\'{\i}}tez}, {Jouvel}, {Seitz}, {Kelson}, \& {Rosati}}]{Coe2013}
{Coe} D. {et~al.}, 2013, \apj, 762, 32

\bibitem[{{Cowie}, {Hu} \& {Songaila}(1995){Cowie}, {Hu}, \&
  {Songaila}}]{Cowie1995}
{Cowie} L.~L., {Hu} E.~M., {Songaila} A., 1995, \aj, 110, 1576

\bibitem[{{Cresci} {et~al}\mbox{.}(2015){Cresci}, {Marconi}, {Zibetti},
  {Risaliti}, {Carniani}, {Mannucci}, {Gallazzi}, {Maiolino}, {Balmaverde},
  {Brusa}, {Capetti}, {Cicone}, {Feruglio}, {Bland-Hawthorn}, {Nagao}, {Oliva},
  {Salvato}, {Sani}, {Tozzi}, {Urrutia}, \& {Venturi}}]{Cresci2015}
{Cresci} G. {et~al.}, 2015, \aap, 582, A63

\bibitem[{{Diego} {et~al}\mbox{.}(2016){Diego}, {Broadhurst}, {Chen}, {Lim},
  {Zitrin}, {Chan}, {Coe}, {Ford}, {Lam}, \& {Zheng}}]{Diego2016}
{Diego} J.~M. {et~al.}, 2016, ArXiv e-prints

\bibitem[{{Diego} {et~al}\mbox{.}(2015{\natexlab{a}}){Diego}, {Broadhurst},
  {Molnar}, {Lam}, \& {Lim}}]{Diego2015a}
{Diego} J.~M., {Broadhurst} T., {Molnar} S.~M., {Lam} D., {Lim} J.,
  2015{\natexlab{a}}, \mnras, 447, 3130

\bibitem[{{Diego} {et~al}\mbox{.}(2015{\natexlab{b}}){Diego}, {Broadhurst},
  {Zitrin}, {Lam}, {Lim}, {Ford}, \& {Zheng}}]{Diego2015b}
{Diego} J.~M., {Broadhurst} T., {Zitrin} A., {Lam} D., {Lim} J., {Ford} H.~C.,
  {Zheng} W., 2015{\natexlab{b}}, \mnras, 451, 3920

\bibitem[{{Diego} {et~al}\mbox{.}(2005a){Diego}, {Protopapas}, {Sandvik}, \&
  {Tegmark}}]{Diego2005}
{Diego} J.~M., {Protopapas} P., {Sandvik} H.~B., {Tegmark} M., 2005a, \mnras,
  360, 477

\bibitem[{{Diego} {et~al}\mbox{.}(2007){Diego}, {Tegmark}, {Protopapas}, \&
  {Sandvik}}]{Diego2007}
{Diego} J.~M., {Tegmark} M., {Protopapas} P., {Sandvik} H.~B., 2007, \mnras,
  375, 958

\bibitem[{{Ebeling}, {White} \& {Rangarajan}(2006){Ebeling}, {White}, \&
  {Rangarajan}}]{Ebeling2006}
{Ebeling} H., {White} D.~A., {Rangarajan} F.~V.~N., 2006, \mnras, 368, 65

\bibitem[{{Elmegreen}, {Elmegreen} \& {Hirst}(2004){Elmegreen}, {Elmegreen}, \&
  {Hirst}}]{Elmegreen2004}
{Elmegreen} D.~M., {Elmegreen} B.~G., {Hirst} A.~C., 2004, \apjl, 604, L21

\bibitem[{{Elmegreen} {et~al}\mbox{.}(2005){Elmegreen}, {Elmegreen}, {Rubin},
  \& {Schaffer}}]{Elmegreen2005}
{Elmegreen} D.~M., {Elmegreen} B.~G., {Rubin} D.~S., {Schaffer} M.~A., 2005,
  \apj, 631, 85

\bibitem[{{Emonts} {et~al}\mbox{.}(2014){Emonts}, {Norris}, {Feain}, {Mao},
  {Ekers}, {Miley}, {Seymour}, {R{\"o}ttgering}, {Villar-Mart{\'{\i}}n},
  {Sadler}, {Carilli}, {Mahony}, {de Breuck}, {Stroe}, {Pentericci}, {van
  Moorsel}, {Drouart}, {Ivison}, {Greve}, {Humphrey}, {Wylezalek}, \&
  {Tadhunter}}]{Emonts2014}
{Emonts} B.~H.~C. {et~al.}, 2014, \mnras, 438, 2898

\bibitem[{{F{\"o}rster Schreiber} {et~al}\mbox{.}(2014){F{\"o}rster Schreiber},
  {Genzel}, {Newman}, {Kurk}, {Lutz}, {Tacconi}, {Wuyts}, {Bandara}, {Burkert},
  {Buschkamp}, {Carollo}, {Cresci}, {Daddi}, {Davies}, {Eisenhauer}, {Hicks},
  {Lang}, {Lilly}, {Mainieri}, {Mancini}, {Naab}, {Peng}, {Renzini}, {Rosario},
  {Shapiro Griffin}, {Shapley}, {Sternberg}, {Tacchella}, {Vergani},
  {Wisnioski}, {Wuyts}, \& {Zamorani}}]{Forster2014}
{F{\"o}rster Schreiber} N.~M. {et~al.}, 2014, \apj, 787, 38

\bibitem[{{Gaibler} {et~al}\mbox{.}(2012){Gaibler}, {Khochfar}, {Krause}, \&
  {Silk}}]{Gaibler2012}
{Gaibler} V., {Khochfar} S., {Krause} M., {Silk} J., 2012, \mnras, 425, 438

\bibitem[{{Genzel} {et~al}\mbox{.}(2014){Genzel}, {F{\"o}rster Schreiber},
  {Rosario}, {Lang}, {Lutz}, {Wisnioski}, {Wuyts}, {Wuyts}, {Bandara},
  {Bender}, {Berta}, {Kurk}, {Mendel}, {Tacconi}, {Wilman}, {Beifiori},
  {Brammer}, {Burkert}, {Buschkamp}, {Chan}, {Carollo}, {Davies}, {Eisenhauer},
  {Fabricius}, {Fossati}, {Kriek}, {Kulkarni}, {Lilly}, {Mancini}, {Momcheva},
  {Naab}, {Nelson}, {Renzini}, {Saglia}, {Sharples}, {Sternberg}, {Tacchella},
  \& {van Dokkum}}]{Genzel2014}
{Genzel} R. {et~al.}, 2014, \apj, 796, 7

\bibitem[{{G{\'o}mez} {et~al}\mbox{.}(2012){G{\'o}mez}, {Valkonen}, {Romer},
  {Lloyd-Davies}, {Verdugo}, {Cantalupo}, {Daub}, {Goldstein}, {Kuo}, {Lange},
  {Lueker}, {Holzapfel}, {Peterson}, {Ruhl}, {Runyan}, {Reichardt}, \&
  {Sabirli}}]{Gomez2012}
{G{\'o}mez} P.~L. {et~al.}, 2012, \aj, 144, 79

\bibitem[{{Gruen} {et~al}\mbox{.}(2013){Gruen}, {Brimioulle}, {Seitz}, {Lee},
  {Young}, {Koppenhoefer}, {Eichner}, {Riffeser}, {Vikram}, {Weidinger}, \&
  {Zenteno}}]{Gruen2013}
{Gruen} D. {et~al.}, 2013, \mnras, 432, 1455

\bibitem[{{Guzzo} {et~al}\mbox{.}(2009){Guzzo}, {Schuecker}, {B{\"o}hringer},
  {Collins}, {Ortiz-Gil}, {de Grandi}, {Edge}, {Neumann}, {Schindler},
  {Altucci}, \& {Shaver}}]{Guzzo2009}
{Guzzo} L. {et~al.}, 2009, \aap, 499, 357

\bibitem[{{Ishigaki} {et~al}\mbox{.}(2015){Ishigaki}, {Kawamata}, {Ouchi},
  {Oguri}, {Shimasaku}, \& {Ono}}]{Ishigaki2015}
{Ishigaki} M., {Kawamata} R., {Ouchi} M., {Oguri} M., {Shimasaku} K., {Ono} Y.,
  2015, \apj, 799, 12

\bibitem[{{Johnson} {et~al}\mbox{.}(2014){Johnson}, {Sharon}, {Bayliss},
  {Gladders}, {Coe}, \& {Ebeling}}]{Johnson2014}
{Johnson} T.~L., {Sharon} K., {Bayliss} M.~B., {Gladders} M.~D., {Coe} D.,
  {Ebeling} H., 2014, \apj, 797, 48

\bibitem[{{Karman} {et~al}\mbox{.}(2015){Karman}, {Caputi}, {Grillo},
  {Balestra}, {Rosati}, {Vanzella}, {Coe}, {Christensen}, {Koekemoer},
  {Kr{\"u}hler}, {Lombardi}, {Mercurio}, {Nonino}, \& {van der
  Wel}}]{Karman2015}
{Karman} W. {et~al.}, 2015, \aap, 574, A11

\bibitem[{{Kelly} {et~al}\mbox{.}(2015){Kelly}, {Rodney}, {Treu}, {Foley},
  {Brammer}, {Schmidt}, {Zitrin}, {Sonnenfeld}, {Strolger}, {Graur},
  {Filippenko}, {Jha}, {Riess}, {Bradac}, {Weiner}, {Scolnic}, {Malkan}, {von
  der Linden}, {Trenti}, {Hjorth}, {Gavazzi}, {Fontana}, {Merten}, {McCully},
  {Jones}, {Postman}, {Dressler}, {Patel}, {Cenko}, {Graham}, \&
  {Tucker}}]{Kelly2016}
{Kelly} P.~L. {et~al.}, 2015, Science, 347, 1123

\bibitem[{{Klamer} {et~al}\mbox{.}(2004){Klamer}, {Ekers}, {Sadler}, \&
  {Hunstead}}]{Klamer2004}
{Klamer} I.~J., {Ekers} R.~D., {Sadler} E.~M., {Hunstead} R.~W., 2004, \apjl,
  612, L97

\bibitem[{{Lotz} {et~al}\mbox{.}(2014){Lotz}, {Mountain}, {Grogin},
  {Koekemoer}, {Capak}, {Mack}, {Coe}, {Barker}, {Adler}, {Avila}, {Anderson},
  {Casertano}, {Christian}, {Gonzaga}, {Ferguson}, {Fruchter}, {Jenkner},
  {Jordan}, {Hammer}, {Hilbert}, {Lawton}, {Lee}, {Lucas}, {MacKenty},
  {Mutchler}, {Ogaz}, {Reid}, {Royle}, {Robberto}, {Sembach}, {Smith}, {Sokol},
  {Surace}, {Taylor}, {Tumlinson}, {Viana}, {Williams}, \&
  {Workman}}]{Lotz2014}
{Lotz} J. {et~al.}, 2014, in American Astronomical Society Meeting Abstracts,
  Vol. 223, American Astronomical Society Meeting Abstracts 223, p. 254.01

\bibitem[{{Meneghetti} {et~al}\mbox{.}(2014){Meneghetti}, {Rasia}, {Vega},
  {Merten}, {Postman}, {Yepes}, {Sembolini}, {Donahue}, {Ettori}, {Umetsu},
  {Balestra}, {Bartelmann}, {Ben{\'{\i}}tez}, {Biviano}, {Bouwens}, {Bradley},
  {Broadhurst}, {Coe}, {Czakon}, {De Petris}, {Ford}, {Giocoli},
  {Gottl{\"o}ber}, {Grillo}, {Infante}, {Jouvel}, {Kelson}, {Koekemoer},
  {Lahav}, {Lemze}, {Medezinski}, {Melchior}, {Mercurio}, {Molino},
  {Moscardini}, {Monna}, {Moustakas}, {Moustakas}, {Nonino}, {Rhodes},
  {Rosati}, {Sayers}, {Seitz}, {Zheng}, \& {Zitrin}}]{Meneghetti2014}
{Meneghetti} M. {et~al.}, 2014, \apj, 797, 34

\bibitem[{{Merten} {et~al}\mbox{.}(2014){Merten}, {Meneghetti}, {Postman},
  {Umetsu}, {Zitrin}, {Medezinski}, {Nonino}, {Koekemoer}, {Melchior}, {Gruen},
  {Moustakas}, {Bartelmann}, {Host}, {Donahue}, {Coe}, {Molino}, {Jouvel},
  {Monna}, {Seitz}, {Czakon}, {Lemze}, {Sayers}, {Balestra}, {Rosati},
  {Ben{\'{\i}}tez}, {Biviano}, {Bouwens}, {Bradley}, {Broadhurst}, {Carrasco},
  {Ford}, {Grillo}, {Infante}, {Kelson}, {Lahav}, {Massey}, {Moustakas},
  {Rasia}, {Rhodes}, {Vega}, \& {Zheng}}]{Merten2014}
{Merten} J. {et~al.}, 2014, ArXiv e-prints

\bibitem[{{Monna} {et~al}\mbox{.}(2014){Monna}, {Seitz}, {Greisel}, {Eichner},
  {Drory}, {Postman}, {Zitrin}, {Coe}, {Halkola}, {Suyu}, {Grillo}, {Rosati},
  {Lemze}, {Balestra}, {Snigula}, {Bradley}, {Umetsu}, {Koekemoer}, {Kuchner},
  {Moustakas}, {Bartelmann}, {Ben{\'{\i}}tez}, {Bouwens}, {Broadhurst},
  {Donahue}, {Ford}, {Host}, {Infante}, {Jimenez-Teja}, {Jouvel}, {Kelson},
  {Lahav}, {Medezinski}, {Melchior}, {Meneghetti}, {Merten}, {Molino},
  {Moustakas}, {Nonino}, \& {Zheng}}]{Monna2014}
{Monna} A. {et~al.}, 2014, \mnras, 438, 1417

\bibitem[{{Oesch} {et~al}\mbox{.}(2014){Oesch}, {Bouwens}, {Illingworth},
  {Labb{\'e}}, {Smit}, {Franx}, {van Dokkum}, {Momcheva}, {Ashby}, {Fazio},
  {Huang}, {Willner}, {Gonzalez}, {Magee}, {Trenti}, {Brammer}, {Skelton}, \&
  {Spitler}}]{Oesch2014}
{Oesch} P.~A. {et~al.}, 2014, \apj, 786, 108

\bibitem[{{Oser} {et~al}\mbox{.}(2012){Oser}, {Naab}, {Ostriker}, \&
  {Johansson}}]{Oser2012}
{Oser} L., {Naab} T., {Ostriker} J.~P., {Johansson} P.~H., 2012, \apj, 744, 63

\bibitem[{{Postman} {et~al}\mbox{.}(2012){Postman}, {Coe}, {Ben{\'{\i}}tez},
  {Bradley}, {Broadhurst}, {Donahue}, {Ford}, {Graur}, {Graves}, {Jouvel},
  {Koekemoer}, {Lemze}, {Medezinski}, {Molino}, {Moustakas}, {Ogaz}, {Riess},
  {Rodney}, {Rosati}, {Umetsu}, {Zheng}, {Zitrin}, {Bartelmann}, {Bouwens},
  {Czakon}, {Golwala}, {Host}, {Infante}, {Jha}, {Jimenez-Teja}, {Kelson},
  {Lahav}, {Lazkoz}, {Maoz}, {McCully}, {Melchior}, {Meneghetti}, {Merten},
  {Moustakas}, {Nonino}, {Patel}, {Reg{\"o}s}, {Sayers}, {Seitz}, \& {Van der
  Wel}}]{Postman2012}
{Postman} M. {et~al.}, 2012, \apjs, 199, 25

\bibitem[{{Rauch} {et~al}\mbox{.}(2013){Rauch}, {Becker}, {Haehnelt},
  {Carswell}, \& {Gauthier}}]{Rauch2013}
{Rauch} M., {Becker} G.~D., {Haehnelt} M.~G., {Carswell} R.~F., {Gauthier}
  J.-R., 2013, \mnras, 431, L68

\bibitem[{{Redlich} {et~al}\mbox{.}(2012){Redlich}, {Bartelmann}, {Waizmann},
  \& {Fedeli}}]{Redlich2012}
{Redlich} M., {Bartelmann} M., {Waizmann} J.-C., {Fedeli} C., 2012, \aap, 547,
  A66

\bibitem[{{Richard} {et~al}\mbox{.}(2014){Richard}, {Jauzac}, {Limousin},
  {Jullo}, {Cl{\'e}ment}, {Ebeling}, {Kneib}, {Atek}, {Natarajan}, {Egami},
  {Livermore}, \& {Bower}}]{Richard2014}
{Richard} J. {et~al.}, 2014, \mnras, 444, 268

\bibitem[{{Salom{\'e}}, {Salom{\'e}} \& {Combes}(2015){Salom{\'e}},
  {Salom{\'e}}, \& {Combes}}]{Salome2015}
{Salom{\'e}} Q., {Salom{\'e}} P., {Combes} F., 2015, \aap, 574, A34

\bibitem[{{Schive}, {Chiueh} \& {Broadhurst}(2014){Schive}, {Chiueh}, \&
  {Broadhurst}}]{Schive2014}
{Schive} H.-Y., {Chiueh} T., {Broadhurst} T., 2014, Nature Physics, 10, 496

\bibitem[{{Schive} {et~al}\mbox{.}(2015){Schive}, {Chiueh}, {Broadhurst}, \&
  {Huang}}]{Schive2015}
{Schive} H.-Y., {Chiueh} T., {Broadhurst} T., {Huang} K.-W., 2015, ArXiv
  e-prints

\bibitem[{{Sendra} {et~al}\mbox{.}(2014){Sendra}, {Diego}, {Broadhurst}, \&
  {Lazkoz}}]{Sendra2014}
{Sendra} I., {Diego} J.~M., {Broadhurst} T., {Lazkoz} R., 2014, \mnras, 437,
  2642

\bibitem[{{Tremblay} {et~al}\mbox{.}(2014){Tremblay}, {Gladders}, {Baum},
  {O'Dea}, {Bayliss}, {Cooke}, {Dahle}, {Davis}, {Florian}, {Rigby}, {Sharon},
  {Soto}, \& {Wuyts}}]{Tremplay2014}
{Tremblay} G.~R. {et~al.}, 2014, \apjl, 790, L26

\bibitem[{{Treu} {et~al}\mbox{.}(2015){Treu}, {Schmidt}, {Brammer}, {Vulcani},
  {Wang}, {Brada{\v c}}, {Dijkstra}, {Dressler}, {Fontana}, {Gavazzi}, {Henry},
  {Hoag}, {Huang}, {Jones}, {Kelly}, {Malkan}, {Mason}, {Pentericci},
  {Poggianti}, {Stiavelli}, {Trenti}, \& {von der Linden}}]{Treu2015}
{Treu} T. {et~al.}, 2015, \apj, 812, 114

\bibitem[{{Umetsu} {et~al}\mbox{.}(2014){Umetsu}, {Medezinski}, {Nonino},
  {Merten}, {Postman}, {Meneghetti}, {Donahue}, {Czakon}, {Molino}, {Seitz},
  {Gruen}, {Lemze}, {Balestra}, {Ben{\'{\i}}tez}, {Biviano}, {Broadhurst},
  {Ford}, {Grillo}, {Koekemoer}, {Melchior}, {Mercurio}, {Moustakas}, {Rosati},
  \& {Zitrin}}]{Umetsu2014}
{Umetsu} K. {et~al.}, 2014, \apj, 795, 163

\bibitem[{{Umetsu} {et~al}\mbox{.}(2015){Umetsu}, {Zitrin}, {Gruen}, {Merten},
  {Donahue}, \& {Postman}}]{Umetsu2015}
{Umetsu} K., {Zitrin} A., {Gruen} D., {Merten} J., {Donahue} M., {Postman} M.,
  2015, ArXiv e-prints

\bibitem[{{van Breugel} {et~al}\mbox{.}(1985){van Breugel}, {Filippenko},
  {Heckman}, \& {Miley}}]{vanBreugel1985}
{van Breugel} W., {Filippenko} A.~V., {Heckman} T., {Miley} G., 1985, \apj,
  293, 83

\bibitem[{{Wagner}, {Umemura} \& {Bicknell}(2013){Wagner}, {Umemura}, \&
  {Bicknell}}]{Wagner2013}
{Wagner} A.~Y., {Umemura} M., {Bicknell} G.~V., 2013, \apjl, 763, L18

\bibitem[{{Zheng} {et~al}\mbox{.}(2014){Zheng}, {Shu}, {Moustakas}, {Zitrin},
  {Ford}, {Huang}, {Broadhurst}, {Molino}, {Diego}, {Infante}, {Bauer},
  {Kelson}, \& {Smit}}]{Zheng2014}
{Zheng} W. {et~al.}, 2014, \apj, 795, 93

\bibitem[{{Zitrin} {et~al}\mbox{.}(2015){Zitrin}, {Fabris}, {Merten},
  {Melchior}, {Meneghetti}, {Koekemoer}, {Coe}, {Maturi}, {Bartelmann},
  {Postman}, {Umetsu}, {Seidel}, {Sendra}, {Broadhurst}, {Balestra}, {Biviano},
  {Grillo}, {Mercurio}, {Nonino}, {Rosati}, {Bradley}, {Carrasco}, {Donahue},
  {Ford}, {Frye}, \& {Moustakas}}]{Zitrin2015}
{Zitrin} A. {et~al.}, 2015, \apj, 801, 44

\bibitem[{{Zitrin} {et~al}\mbox{.}(2014){Zitrin}, {Zheng}, {Broadhurst},
  {Moustakas}, {Lam}, {Shu}, {Huang}, {Diego}, {Ford}, {Lim}, {Bauer},
  {Infante}, {Kelson}, \& {Molino}}]{Zitrin2014}
{Zitrin} A. {et~al.}, 2014, \apjl, 793, L12

\end{thebibliography}


\appendix

\section{Compilation of arc positions}

    \begin{table*}
    \begin{minipage}{165mm}                                               
    \caption{Full strong lensing data set. The first column shows system ID following the original notation 
             of Richard et al. (2014) and Johnson et al. (2014). New systems are marked with an $*$ in the Notes column.
             Systems marked with $**$ are found after the derivation of the lens models but not used as constraints. 
             A C in the Notes column shows the position of a counterimage candidate but not used in the analysis. An L 
             in the Notes column marks a long arc which may or may not be multiply lensed. 
             System 52 is found after using the new redshift quoted in \citep{Caminha2015a}.  
             The second and third columns show the coordinates of each arclet. 
             Column 4 includes the redshifts (and references for the spectroscopic redshifts, 
             1=\citep{Balestra2013}, 2=\citep{Boone2013}, 3=\citep{Richard2014}, 4=\citep{Karman2015},
             5=\citep{Johnson2014},  6=\citep{Caminha2015a}, 7=\citep{Monna2014}, 8=GLASS,\citep{Treu2015} ) used in our study.  
             Spectroscopic redshifts are marked in bold face. The remaining redshifts are estimated from the lens model 
             derived using the systems with spectroscopic redshifts }
 \label{tab1}
 \begin{tabular}{ccccc}   
  ID     &  RAJ2000(h:m:s)  & DECJ2000(d:m:s)  &    z    &       Notes      \\
 1.1     &  22:48:46.668    &  -44:31:37.21  & {\bf 1.229$^{1,3,5,6}$} &      \\        
 1.2     &  22:48:47.008    &  -44:31:44.22  &    "      &                  \\     
 1.3     &  22:48:44.741    &  -44:31:16.33  &    "      &                  \\     
 2.1     &  22:48:46.250    &  -44:31:52.28  & {\bf 1.261$^{1,3,5,6}$} &      \\        
 2.2     &  22:48:46.110    &  -44:31:47.39  &    "      &                  \\     
 2.3     &  22:48:43.167    &  -44:31:17.62  &    "      &                  \\     
 3.1     &  22:48:46.930    &  -44:31:55.70  &     1.7   &                  \\     
 3.2     &  22:48:46.540    &  -44:31:43.43  &    "      &                  \\     
 4.1     &  22:48:46.490    &  -44:31:48.58  &     1.2   &                  \\     
 4.2     &  22:48:46.400    &  -44:31:45.91  &    "      &                  \\     
 5.1     &  22:48:43.010    &  -44:31:24.92  & {\bf 1.398$^{1,3,5,6}$} &      \\        
 5.2     &  22:48:45.080    &  -44:31:38.32  &    "      &                  \\     
 5.3     &  22:48:46.360    &  -44:32:11.51  &    "      &                  \\     
 6.1     &  22:48:41.820    &  -44:31:41.99  & {\bf 1.428$^{1,3,4,5}$} &     \\       
 6.2     &  22:48:42.200    &  -44:31:57.14  &    "      &                  \\     
 6.3     &  22:48:45.225    &  -44:32:23.98  &    "      &                  \\     
 7.1     &  22:48:40.650    &  -44:31:38.10  &  {\bf  1.837$^{6}$}    &      \\       
 7.2     &  22:48:41.820    &  -44:32:13.60  &    "      &                  \\     
 7.3     &  22:48:43.640    &  -44:32:25.80  &    "      &                  \\     
 8.1     &  22:48:40.310    &  -44:31:34.32  &     2.8   &                  \\     
 8.2     &  22:48:41.910    &  -44:32:18.20  &    "      &                  \\     
 8.3     &  22:48:43.390    &  -44:32:27.17  &    "      &                  \\     
 9.1     &  22:48:40.270    &  -44:31:34.61  &     2.8   &                  \\     
 9.2     &  22:48:41.950    &  -44:32:19.00  &    "      &                  \\     
 9.3     &  22:48:43.270    &  -44:32:26.92  &    "      &                  \\     
 10.1    &  22:48:45.657    &  -44:31:47.15  & {\bf 0.73$^{6,8}$}     &        \\     
 10.2    &  22:48:45.492    &  -44:31:43.83  &    "      &           (*)    \\     
 10.3    &  22:48:44.380    &  -44:31:31.71  &    "      &           (*)    \\     
 11.1    &  22:48:42.010    &  -44:32:27.71  & {\bf 3.116$^{1,4,5,6}$} &      \\      
 11.2    &  22:48:41.560    &  -44:32:23.93  &    "      &                  \\     
 11.3    &  22:48:39.733    &  -44:31:46.31  &    "      &                  \\     
 12.1    &  22:48:45.370    &  -44:31:48.18  & {\bf 6.112$^{1,2,4,7}$} &      \\     
 12.2    &  22:48:43.450    &  -44:32:04.63  &    "      &                  \\     
 12.3    &  22:48:45.810    &  -44:32:14.89  &    "      &                  \\     
 12.4    &  22:48:41.110    &  -44:31:11.32  &    "      &                  \\     
 13.1    &  22:48:43.572    &  -44:32:21.75  & {\bf  4.113$^{4}$}    &       \\     
 13.2    &  22:48:42.993    &  -44:32:19.24  &    "      &                  \\     
 14.1    &  22:48:42.920    &  -44:32:09.13  & {\bf  3.118$^{6}$}    &       \\     
 14.2    &  22:48:44.980    &  -44:32:19.28  &    "      &                  \\     
 14.3    &  22:48:40.960    &  -44:31:19.52  &    "      &                  \\     
 15.1    &  22:48:46.010    &  -44:31:49.87  &    2.5    &                  \\     
 15.2    &  22:48:46.210    &  -44:32:03.91  &    "      &                  \\     
 15.3    &  22:48:42.220    &  -44:31:10.74  &    "      &                  \\     
 16.1    &  22:48:39.900    &  -44:32:01.14  &    3.1    &                  \\     
 16.2    &  22:48:40.030    &  -44:32:05.75  &    "      &                  \\     
 16.3    &  22:48:42.680    &  -44:32:35.05  &    "      &                  \\     
 17.1    &  22:48:44.600    &  -44:32:19.86  &    3.1    &                  \\     
 17.2    &  22:48:42.920    &  -44:32:12.23  &    "      &                  \\     
 17.3    &  22:48:40.750    &  -44:31:19.12  &    "      &                  \\     
 18.1    &  22:48:41.320    &  -44:32:11.83  &    3.5    &                  \\     
 18.2    &  22:48:44.350    &  -44:32:31.42  &    "      &                  \\     
 19.1    &  22:48:43.205    &  -44:32:18.35  & {\bf  1.035$^{4}$}   &      \\     
 19.2    &  22:48:42.130    &  -44:32:09.38  &    "      &                 \\     
 19.3    &  22:48:41.260    &  -44:31:48.90  &    "      &                 \\     
 20.1    &  22:48:51.830    &  -44:31:09.94  &    2.0    &                 \\     
 20.2    &  22:48:51.700    &  -44:31:08.78  &    "      &                 \\     
 20.3    &  22:48:51.390    &  -44:31:00.63  &    "      &                 \\      
 21.1    &  22:48:44.877    &  -44:31:38.70  &    0.75   &           (*) \\
 21.2    &  22:48:44.450    &  -44:31:34.82  &    "      &           (*) \\     
 21.3    &  22:48:44.649    &  -44:31:36.63  &    "      &           (*) \\          
 21.4    &  22:48:45.685    &  -44:31:53.65  &    "      &           (*) \\     
 \end{tabular}
 \end{minipage}
\end{table*}
\setcounter{table}{0}
    \begin{table*}
    \begin{minipage}{165mm}                                               
    \caption{cont.}
 \begin{tabular}{ccccc}    
  ID     &  RAJ2000(h:m:s)  & DECJ2000(d:m:s)  &    z    &    Notes      \\
 22.1    &  22:48:47.710    &  -44:31:10.64  &    3      &           (*,L) \\     
 22.2    &  22:48:48.116    &  -44:31:16.42  &    "      &           (*,L) \\          
 22.3    &  22:48:48.423    &  -44:31:21.09  &    "      &           (*,L) \\     
 23.1    &  22:48:41.019    &  -44:31:46.55  &    3.5    &           (*) \\      
 23.2    &  22:48:41.168    &  -44:31:56.51  &    "      &           (*) \\     
 24.1    &  22:48:50.870    &  -44:31:18.83  &    3.0    &           (*,L) \\      
 24.2    &  22:48:50.748    &  -44:31:16.58  &    "      &           (*,L) \\     
 24.3    &  22:48:50.577    &  -44:31:13.63  &    "      &           (*,L) \\     
 25.1    &  22:48:51.644    &  -44:31:09.23  &    3.0    &           (*) \\     
 25.2    &  22:48:51.545    &  -44:31:07.89  &    "      &           (*) \\     
 25.3    &  22:48:51.461    &  -44:31:04.93  &    "      &           (*) \\     
 26.1    &  22:48:49.243    &  -44:31:28.99  &    3.0    &           (*,L) \\     
 26.2    &  22:48:48.980    &  -44:31:24.66  &    "      &           (*,L) \\     
 28.1    &  22:48:42.084    &  -44:32:23.05  &    1.3    &           (*) \\     
 28.2    &  22:48:41.875    &  -44:32:21.26  &    "      &           (*) \\   
 28.3    &  22:48:40.532    &  -44:31:57.19  &    "      &           (*,C) \\   
 31.1    &  22:48:47.655    &  -44:31:14.90  &    3.5    &           (*) \\     
 31.2    &  22:48:47.353    &  -44:31:11.17  &    "      &           (*) \\     
 32.1    &  22:48:43.255    &  -44:32:21.35  &    2.0    &           (*) \\     
 32.2    &  22:48:42.945    &  -44:32:19.90  &    "      &           (*) \\     
 34.1    &  22:48:45.854    &  -44:31:23.97  &    1.9    &           (*,L) \\     
 34.2    &  22:48:45.723    &  -44:31:22.65  &    "      &           (*,L) \\     
 34.3    &  22:48:45.497    &  -44:31:20.80  &    "      &           (*,L) \\     
 34.4    &  22:48:45.298    &  -44:31:19.45  &    "      &           (*,L) \\     
 41.1    &  22:48:37.141    &  -44:32:22.39  &    2.0    &           (*) \\     
 41.2    &  22:48:37.098    &  -44:32:22.33  &    "      &           (*) \\     
 41.3    &  22:48:37.030    &  -44:32:19.92  &    "      &           (*) \\     
 43.1    &  22:48:42.548    &  -44:32:26.53  &    2.0    &           (*) \\     
 43.2    &  22:48:41.191    &  -44:32:13.82  &    "      &           (*) \\     
 43.3    &  22:48:40.338    &  -44:31:53.33  &    "      &           (*) \\     
 44.1    &  22:48:47.607    &  -44:32:08.71  &     2.0   &           (*) \\     
 44.2    &  22:48:46.186    &  -44:31:30.25  &    "      &           (*) \\     
 44.3    &  22:48:43.561    &  -44:31:12.93  &    "      &           (*) \\     
 45.1    &  22:48:47.733    &  -44:32:05.16  &    2.0    &           (*) \\     
 45.2    &  22:48:46.388    &  -44:31:29.43  &    "      &           (*) \\     
 45.3    &  22:48:43.944    &  -44:31:11.60  &    "      &           (*) \\     
 46.1     &  22:48:46.007    &  -44:32:12.08  &   1.28   &           (**) \\     
 46.2     &  22:48:44.782    &  -44:31:41.82  &    "     &           (**) \\     
 46.3     &  22:48:42.817    &  -44:31:27.94  &    "     &           (**) \\     
 47.1     &  22:48:46.332    &  -44:32:07.91  &   2.9    &           (**) \\     
 47.2     &  22:48:45.901    &  -44:31:45.20  &    "     &           (**) \\     
 47.2     &  22:48:41.872    &  -44:31:12.30  &    "     &           (**) \\     
 48.1     &  22:48:51.998    &  -44:30:59.23  &    3     &           (**) \\     
 48.2     &  22:48:52.306    &  -44:31:07.48  &    "     &           (**) \\     
 48.3     &  22:48:52.281    &  -44:31:06.06  &    "     &           (**) \\     
 49.1     &  22:48:46.061    &  -44:32:12.56  &   1.2    &           (**) \\     
 49.2     &  22:48:44.703    &  -44:31:41.05  &    "     &           (**) \\     
 49.3     &  22:48:42.707    &  -44:31:28.62  &    "     &           (**) \\     
 50.1     &  22:48:44.895    &  -44:31:44.79  &   1.7    &           (**) \\     
 50.2     &  22:48:43.445    &  -44:31:56.72  &    "     &           (**) \\     
 50.3     &  22:48:42.155    &  -44:31:23.48  &    "     &           (**) \\     
 51.1     &  22:48:47.388    &  -44:31:50.28  &   1.9    &           (**) \\     
 51.2     &  22:48:47.133    &  -44:31:42.21  &    "     &           (**) \\     
 51.3     &  22:48:43.959    &  -44:31:08.70  &    "     &           (**) \\     
 52.1     &  22:48:41.741    &  -44:32:28.46  & {\bf 3.228$^{4,6}$}  &   (**)  \\
 52.2     &  22:48:39.614    &  -44:31:51.80  &    "     &           (**) \\
 14.1.a   &  22:48:43.109    &  -44:32:08.60  &   3.1    &           (**) \\     
 14.1.b   &  22:48:43.197    &  -44:32:08.17  &    "     &           (**) \\     
 14.1.c   &  22:48:42.779    &  -44:32:09.61  &    "     &           (**) \\     
 14.1.d   &  22:48:42.782    &  -44:32:08.96  &    "     &           (**) \\     
 14.2.a   &  22:48:45.065    &  -44:32:17.81  &    "     &           (**) \\     
 14.2.b   &  22:48:45.085    &  -44:32:16.95  &    "     &           (**) \\     
 14.2.c   &  22:48:44.914    &  -44:32:20.55  &    "     &           (**) \\     
 14.2.d   &  22:48:45.041    &  -44:32:20.40  &    "     &           (**) \\     
 14.3.a   &  22:48:40.999    &  -44:31:18.09  &    "     &           (**) \\     
 14.3.b   &  22:48:41.025    &  -44:31:17.38  &    "     &           (**) \\     
 14.3.c   &  22:48:40.915    &  -44:31:20.87  &    "     &           (**) \\     
 14.3.d   &  22:48:40.945    &  -44:31:20.58  &    "     &           (**) \\     
 \end{tabular}
 \end{minipage}
\end{table*}

\end{document}